\documentclass[pre,aps,floats,twocolumn,superscriptaddress,floatfix]{revtex4}
\usepackage{graphicx,amssymb,amsmath,ifthen,rotating}
\usepackage[active]{srcltx}
\begin{document}
\title{Mechanical response to tension and torque of molecular chains via statistically interacting particles associated with extension, contraction, twist, and supercoiling}  
\author{Aaron C. Meyer}
\affiliation{
  Department of Physics,
  University of Rhode Island,
  Kingston RI 02881, USA}
  \author{Michael Karbach}
\affiliation{
Fachgruppe Physik,
  Bergische Universit{\"{a}}t Wuppertal,
  D-42097 Wuppertal, Germany}
        \author{Ping Lu}
\affiliation{
Department of Physics,
  Stetson University,
  DeLand, FL 32723, USA} 
\author{Gerhard M{\"{u}}ller}
\affiliation{
  Department of Physics,
  University of Rhode Island,
  Kingston RI 02881, USA}

\begin{abstract}
A methodology for the statistical mechanical analysis of polymeric chains under tension introduced previously is extended to include torque.
The response of individual bonds between monomers or of entire groups of monomers to a combination of tension and torque involves, in the framework of this method of analysis,  the (thermal or mechanical) activation of a specific mix of statistically interacting particles carrying quanta of extension or contraction and quanta of twist or supercoiling.
The methodology, which is elucidated in applications of increasing complexity, is capable of describing the conversion between twist chirality and plectonemic chirality in quasistatic processes.
The control variables are force or extension and torque or linkage (a combination of twist and writhe).
The versatility of this approach is demonstrated in two applications relevant and promising for double-stranded DNA under controlled tension and torque.
One application describes conformational transformations between (native) B-DNA, (underwound) S-DNA, and (overwound) P-DNA in accord with experimental data.
The other application describes how the conversion between a twisted chain and a supercoiled chain accommodates variations of linkage and excess length in a buckling transition.
\end{abstract}
\maketitle

%
\section{Introduction}\label{sec:intro}
%
This work investigates the statistical mechanics of molecular chains with (internal and external) torsional constraints.
The bonds between monomers are not rotatable, neither freely nor across periodic energy barriers. 
In consequence, the molecular chain responds to torque by a buildup of torsional elastic energy, which is necessarily coupled to elastic energy associated with elongation or contraction.
Hence torque and tension both vary, in general, when the molecular chain is being twisted or stretched.
Research on double-stranded (ds) DNA is the most notable field of applications by far  \cite{MS94, MS95, SAB+96, Mark97, Mark98, ABLC98, Smit98, SABC98, HYZ99, SAB+99, LRS+99, BM00, ZZY00, SABC00, SACB00, SLCM01, BBS03, BSG+03, Neuk04, LJL+06, Mark07, FDS+08, DFS+09, SW09, WRC09, NR11, GLO+11, OK11, OK12, MN13,ELS13}.

Experimentally, the ends of the chain are mounted to the measuring device such that the total angle of twist plus writhe (named linkage) is either held constant or is controllably changed.
In some instances, the application of torque (or tension) nucleates conformational transformations.
One conformation then grows at the expense of another.
In other instances, a coupled variation of torque and tension, initiates and grows supercoiling, converting twist into writhe and changing the excess length simultaneously in the process.

In this setup, it is necessary to consider a mechanical agent which simultaneously exerts tension $J$ and torque $\tau$ in some combination.
The molecular chain responds with a combination of extension/contraction length $L$ and twist/writhe angle $\phi$.
The statistical mechanical analysis, therefore, deals with three pairs of conjugate thermodynamic variables, $(J,L)$, $(\tau,\phi)$, and $(T,S)$, where the last pair are temperature and entropy.

Our method of analysis  leads, by default, to a Gibbs free energy with natural independent variables $T,J,\tau$, from which the thermal and mechanical responses are derived via derivatives.
Thermodynamic relations are readily inverted to produce relations amenable to direct comparisons with experimental data.
This method of analysis was previously introduced in some detail for applications to molecular chains under tension without torsional constraints \cite{mct1}.

Recent advances in single-molecule biophysics, specifically the enormous progress achieved in the experimental investigation of DNA double helices subject to controllable tension and torque, are in need of ever more versatile theoretical approaches for the interpretation of new data.
This paper offers a contribution with a demonstration of its merits to satisfy that need.
It comprises the natural extension of a general methodology known as fractional exclusion statistics (FES) \cite{Hald91a, Wu94, Isak94, Anghel, NA14, LVP+08, copic, picnnn, pichs}. 

The approach has already proven its usefulness in numerous quantum and classical applications \cite{LMK09, PMK07, sivp, GKLM13, janac2, cohetra}, notably for ds-DNA subjected to torsionally unconstrained stretching \cite{mct1}.
The results of this approach aligned with experimental data of the force-extension characteristic across regimes of entropic elasticity (thermal umbending), enthalpic elasticity (beyond contour length), and an overstretching transition.

The technical aspects of the statistical mechanical analysis, which is firmly grounded in quasistatic processes, are briefly reviewed in Sec.~\ref{sec:metho} with emphasis on those aspects that are in need of eloborations beyond the account given in Ref.~\cite{mct1}.
One elementary application, which illustrates the coupling between the two pairs $(J,L)$ and $(\tau,\phi)$ of conjugate thermodynamic variables in the context of torsionally constrained twisting and stretching, is worked out in Sec.~\ref{sec:twis-cont}.

The focus of Sec.~\ref{sec:BSP-DNA} is on transitions between three well established conformations of ds-DNA. 
We infer from a single partition function the empirically found phase diagram with three well-defined boundaries between (native) B-DNA, (underwound) S-DNA, and (overwound) P-DNA.
The coexistence between twisted chain and supercoiled chain described within the same framework is the topic of Sec.~\ref{sec:sup-coi}.
The distinction is made between high-tension supercoils (rope variety) and  low-tension supercoils (garden-hose variety), analyzed in Secs.~\ref{sec:sup-coi-ht} and \ref{sec:sup-coi-lt}, respectively.
The former is set up for the demonstration in Sec.~\ref{sec:DNA-plect} of a buckling transition. 

%
\section{Methodology}\label{sec:metho}
%
Quasistatic processes operate within the realm of equilibrium statistical mechanics. 
The response of a thermodynamic system (here a molecular chain embedded in a fluid) to agents of change (here sources of tension and torque) is described by thermodynamic relations derived from a partition function.
The partition function is a sum over microstates of terms weighted according to energies.
Microstates are characterized by their quasiparticle content.
Quasiparticles are (thermally or mechanically) activated from some reference state. 

For a system of statistically interacting particles from species $m=1,\ldots,M$, the partition function, 
\begin{equation}\label{eq:1} 
Z=\sum_{\{N_m\}}W(\{N_m\})e^{-\beta E(\{N_m\})},\quad \beta\doteq\frac{1}{k_\mathrm{B}T},
\end{equation}
depends on the multiplicity of microstates,
\begin{subequations}\label{eq:2} 
\begin{align}\label{eq:2a} 
W(\{N_m\}) &=\prod_{m=1}^M\left(\begin{array}{c}
d_m+N_m-1 \\ N_m\end{array}\right), \\ \label{eq:2b} 
 d_m &=A_m-\sum_{m'=1}^M g_{mm'}(N_{m'}-\delta_{mm'}),
\end{align}
\end{subequations}
and their energies,
\begin{equation}\label{eq:3} 
E(\{N_m\})=E_{\mathrm{pv}}+\sum_{m=1}^M N_m\epsilon_m,
\end{equation}
where $E_\mathrm{pv}$ is the energy of the reference state (pseudo-vacuum) and the $\epsilon_m$ are the particle activation energies.
Multiplicity $W$ and energy $E$ are functions of particle content $\{N_m\}$.
The statistical interactions between particles are encoded in arrays of (non-negative, rational) capacity constants $A_m$ and (rational) statistical interaction coefficients $g_{mm'}$.

The keystone in this scheme is the generalized Pauli principle, of which (\ref{eq:2b}) is an integrated version.
It was proposed and first used by Haldane \cite{Hald91a}.
The evaluation of (\ref{eq:1}) was investigated by Wu \cite{Wu94}, Isakov \cite{Isak94}, Anghel \cite{Anghel}, and others \cite{LMK09, PMK07, sivp} at various levels of generality.
For a macroscopic system, we can write the partition function in the form,
\begin{equation}\label{eq:4} 
Z=\prod_{m=1}^M\big(1+w_m^{-1}\big)^{A_m},
\end{equation}
where the (real, positive) $w_m$ are solutions of the coupled nonlinear algebraic equations,
\begin{equation}\label{eq:5} 
e^{\beta\epsilon_m}=(1+w_m)\prod_{m'=1}^M \big(1+w_{m'}^{-1}\big)^{-g_{m'm}}.
\end{equation}
The average numbers $\langle N_m\rangle$ of particles from all species are the solutions, for given $\{w_m\}$, of the coupled linear equations,
\begin{equation}\label{eq:6} 
w_m\langle N_m\rangle+\sum_{m'=1}^Mg_{mm'}\langle N_{m'}\rangle =A_m.
\end{equation}

In this work, all particles carry quanta of length and angle. 
Their activation energies are of the general form,
\begin{equation}\label{eq:7} 
\epsilon_m=\gamma_m-JL_m-\tau\phi_m.
\end{equation}
The quantum of length $L_m$ is positive for extension particles and negative for contraction particles.
Likewise, the quantum of angle $\phi_m$ can be positive or negative. 
In DNA applications, particles with positive $\phi_m$ overwind the double helix and particles with negative $\phi_m$ underwind it when activated.
The angle $\phi_m$ represents twist in some particle species and writhe in others.
Twist particles and writhe particles typically have quite different length quanta $L_m$.
The role of the energy constant $\gamma_m$ varies between species and applications.

We infer from the partition function (\ref{eq:4}) the Gibbs free energy,
\begin{equation}\label{eq:8} 
G(T,J,\tau,N)=-k_\mathrm{B}T\ln Z,
\end{equation}
where the dependence on $T$ comes from $\beta$ in (\ref{eq:5}), the dependence on $J$, $\tau$ from (\ref{eq:7}) via (\ref{eq:5}) and the dependence on $N$ (the number of bonds) is hidden in the capacity constant $A_m$ of one or several particle species in a way that guarantees thermodynamic extensivity of $G$.
The quantities of primary interest here are entropy, excess length, and linkage, obtained via partial derivatives,
\begin{equation}\label{eq:9} 
S =-\frac{\partial G}{\partial T},\quad
\langle L\rangle=-\frac{\partial G}{\partial J},\quad
\langle\phi\rangle=-\frac{\partial G}{\partial \tau}.
\end{equation}
These thermodynamic functions can also be inferred from particle population averages via Eqs.~(\ref{eq:6}) \cite{mct1,Isak94,sivp}:
\begin{subequations}\label{eq:17}
\begin{align}\label{eq:17a}
&S = k_\mathrm{B}\sum_{m=1}^M
\Big[\big(\langle N_m\rangle+{Y}_m\big)\ln\big(\langle N_m\rangle+{Y}_m\big) \nonumber \\
&\hspace{20mm}-\langle N_m\rangle\ln \langle N_m\rangle -{Y}_m\ln {Y}_m\Big], \\ \label{eq:17b}
&{Y}_m \doteq {A}_m-\sum_{m'=1}^Mg_{mm'} \langle N_{m'}\rangle,
\end{align}
\end{subequations}
\begin{equation}\label{eq:16} 
 \langle L\rangle=\sum_{m=1}^ML_m\langle N_m\rangle,\quad 
 \langle\phi\rangle=\sum_{m=1}^M\phi_m\langle N_m\rangle.
\end{equation}

In Ref.~\cite{mct1} we explained the different categories of particles: compact particles, which exist side by side, and nested particles, which form hierarchical structures.
Particles at level 1 modify individual bonds and particles at level 2 modify entire segments of the molecular chain. 
We also worked out general solutions for several sets of particles. 
None of this will be reiterated here.
Instead we will comment on all essentials in the context of each application with pointers to prior work.

%
\section{Twist contraction}\label{sec:twis-cont}
%
The first application pertains to an idealized double-stranded molecular chain whose reference state is a ladder with no native helical structure.
The applied tension $J$ is assumed to remain below the threshold of significant contour elasticity such as discussed in Sec. III of \cite{mct1}. 
If a torque $\tau$ is applied the ladder conformation responds by a combination of twist and contraction.

We model this response with a system of two species of particles that carry quanta of twist angle and contraction length.
The combinatorial specifications of these particles are compiled in Table~\ref{tab:t1}. 
They are level-1 compacts as introduced in Sec.~II.E of Ref.~\cite{mct1} along with a general solution.
The particle activation energies have the general form (\ref{eq:7}), which we rewrite as
\begin{equation}\label{eq:10} 
\epsilon_\pm=\gamma_\mathrm{t}+JL_\mathrm{t}\mp\tau\phi_\mathrm{t}.
\end{equation}
The three energetic specifications are an elastic energy constant $\gamma_\mathrm{t}>0$, a quantum of contraction length $L_\mathrm{t}>0$, and a quantum of twist angle $\phi_\mathrm{t}>0$.
The two particle species account for the two senses of torque and twist, both of which are associated with a contraction.

\begin{table}[htb]
  \caption{Capacity constants $A_m$ and statistical interaction coefficients $g_{mm'}$ for the two species of level-1 compacts.}\label{tab:t1}
\begin{center}
\begin{tabular}{c|c} \hline\hline
$m$~~ & ~~$A_m$  \\ \hline
$+$~~ & ~~$N-1$ \\
$-$~~ & ~~$N-1$ \\ 
\hline\hline 
\end{tabular} \hspace{5mm}
\begin{tabular}{c|rr} \hline\hline 
$g_{mm'}$~ & ~~$+$ & ~~$-$  \\ \hline 
$+$ & $1$ & $1$ \\ 
$-$ & $0$ & $1$ \\ 
 \hline\hline
\end{tabular}
\end{center}
\end{table}

If a positive torque $\tau$ is applied at constant tension $J$, the activation energy $\epsilon_+$ decreases, whereas $\epsilon_-$ increases.
The effect is a positive twist and a contraction. 
Likewise, a negative torque favors the activation of particles with activation energies $\epsilon_-$, which produces a negative twist and a contraction again.
On the other hand, if we increase $J$ at constant (positive or negative) $\tau$, then both activation energies $\epsilon_\pm$ increase.
Twist particles from both species are gradually frozen out.
As the chain untwists, its contraction diminishes.

The Gibbs free energy per bond of a long chain, $\bar{G}(T,J,\tau)$, inferred from the solution of Eqs.~(\ref{eq:5}),
\begin{equation}\label{eq:11} 
w_+=e^{\beta\epsilon_+},\quad
w_-=e^{\beta\epsilon_-}\big(1+e^{-\beta\epsilon_+}\big)
\end{equation}
via (\ref{eq:4}) and (\ref{eq:8}), becomes
\begin{equation}\label{eq:12} 
\bar{G}\doteq\lim_{N\to\infty}\frac{G}{N}=-k_\mathrm{B}T\ln\Big(1+e^{-\beta\epsilon_+}+e^{-\beta\epsilon_-}\Big).
\end{equation}
The first partial derivatives (\ref{eq:9}), 
\begin{subequations}\label{eq:13} 
\begin{align}\label{eq:13a} 
\bar{S}\doteq\lim_{N\to\infty}\frac{S}{Nk_\mathrm{B}} &=\Bigg[\ln\Big(1+e^{-\beta\epsilon_+}+e^{-\beta\epsilon_-}\Big)  \nonumber \\
&\hspace{0mm} \hspace{5mm}+\frac{\beta\epsilon_+ e^{-\beta\epsilon_+}+\beta\epsilon_- 
e^{-\beta\epsilon_-}}{1+e^{-\beta\epsilon_+}+e^{-\beta\epsilon_-}}\Bigg],
\end{align}
\begin{equation}\label{eq:13b} 
\bar{L}\doteq\lim_{N\to\infty}\frac{\langle L\rangle}{N}
=-L_\mathrm{t}\,\frac{e^{-\beta\epsilon_+}
+e^{-\beta\epsilon_-}}{1+e^{-\beta\epsilon_+}+e^{-\beta\epsilon_-}},
\end{equation}
\begin{equation}\label{eq:13c} 
\bar{\phi}\doteq\lim_{N\to\infty}\frac{\langle\phi\rangle}{N}
= \phi_\mathrm{t}\,\frac{e^{-\beta\epsilon_+}-e^{-\beta\epsilon_-}}
{1+e^{-\beta\epsilon_+}+e^{-\beta\epsilon_-}},
\end{equation}
\end{subequations}
represent entropy, (negative) extension length, and twist angle, respectively.

\begin{figure}[b]
  \begin{center}
\includegraphics[width=40mm]{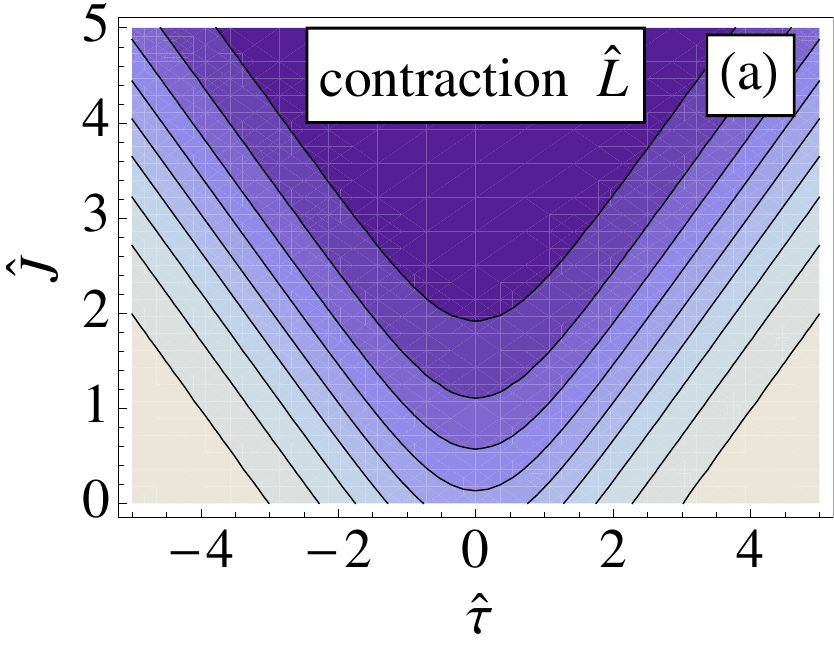}\hspace*{3mm}\includegraphics[width=40mm]{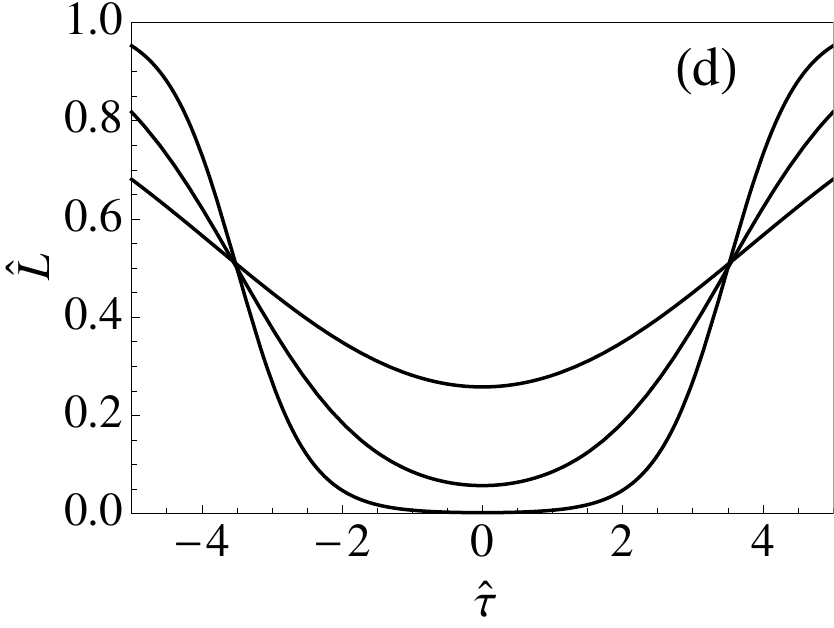}
\includegraphics[width=40mm]{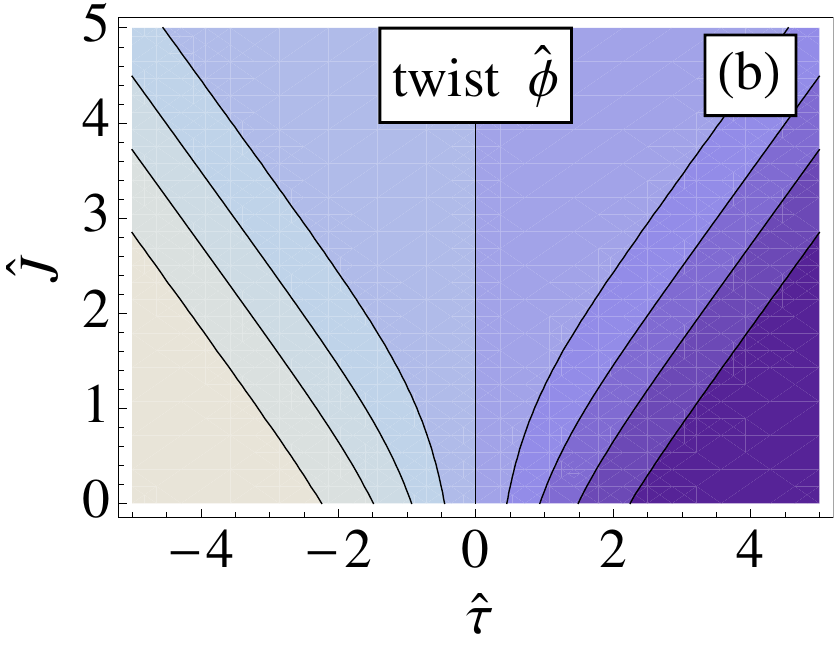}\hspace*{3mm}\includegraphics[width=40mm]{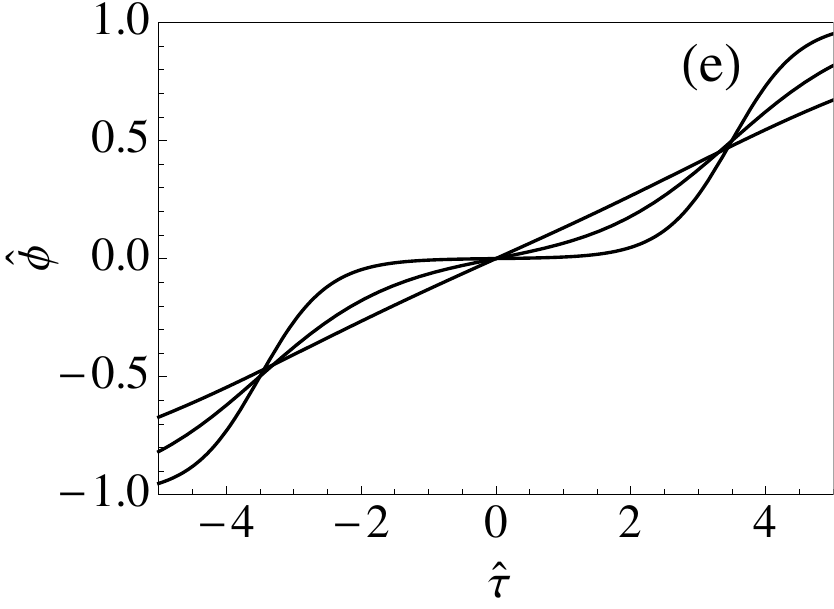}
\includegraphics[width=40mm]{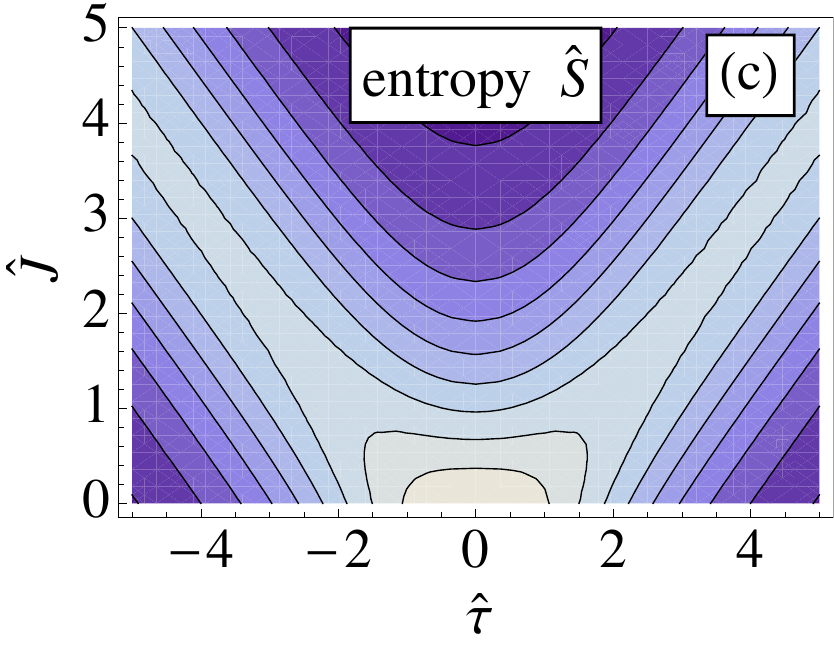}\hspace*{3mm}\includegraphics[width=40mm]{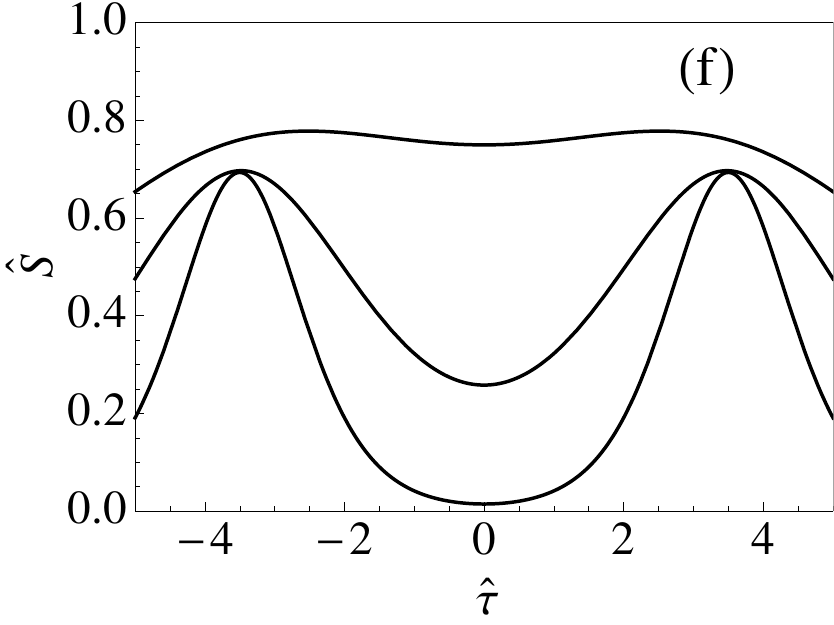}
\end{center}
\caption{(a) Contraction distance $\hat{L}$, (b) twist angle $\hat{\phi}$, and (c) entropy $\hat{S}$, all versus tension $\hat{J}$ and torque $\hat{\tau}$ at constant temperature $\hat{T}=1$. The nine contour lines from dark to bright in each panel are at (a) $\hat{L}=0.098, \ldots, 0.882$, (b) $\hat{\phi}=-0.76, \ldots, 0.76$, and (c) $\hat{S}=0.097, \ldots, 0.873$. 
Panels (d)-(f) show the same quantities plotted versus $\hat{\tau}$ at constant $\hat{J}=2.5$ and $\hat{T}=0.5,1,2$.}
  \label{fig:figure5}
\end{figure}

The contour plots in Fig.~\ref{fig:figure5} visualize (in scaled units) the dependence of contraction distance $\hat{L}\doteq|\bar{L}|/L_\mathrm{t}$, twist angle $\hat{\phi}\doteq\bar{\phi}/\phi_\mathrm{t}$, and entropy $\hat{S}\doteq\bar{S}/k_\mathrm{B}$ on tension $\hat{J}\doteq JL_\mathrm{t}/\gamma_\mathrm{t}$ and torque $\hat{\tau}\doteq \tau\phi_\mathrm{t}/\gamma_\mathrm{t}$, the two mechanical control variables, all at constant temperature $\hat{T}\doteq k_\mathrm{B}T/\gamma_\mathrm{t}$. 
The landscape in panel (a) describes how the system contracts when the torque increases at constant tension and how the system expands when the tension grows at constant torque.

The twist angle responds to a torque of increasing magnitude as shown in panel (b).
The response is antisymmetric and stronger at low tension than at high tension.
If we apply a torque of significant strength in positive direction, then only twist particles with activation energies $\epsilon_+$ attain significant populations. 

The entropy landscape  of panel (c) is more complicated. 
The entropy is low at strong tension and weak torque because twist particles are mostly frozen out owing to their high activation energies. 
The entropy is also low at strong torques and low tension.
Here the system is saturated with one or the other species of twist particles, which have negative activation energies.
Elsewhere, the macrostate of the system is more strongly fluctuating, which enhances the entropy.

Varying the scaled temperature shifts the balance between the quanta of elastic energy carried by the particles and the ambient thermal fluctuations.
The effects are illustrated in Figs.~\ref{fig:figure5}(d)-(f).
For all three quantities, the slopes characterizing the landscapes become steeper as $\hat{T}$ is lowered. 
Emerging are two steps for $\hat{L}$, a terrace with three levels for $\hat{\phi}$, and two narrow ridges for $\hat{S}$.

%
\section{From B-DNA to S-DNA and P-DNA}\label{sec:BSP-DNA}
%
Here we generalize the previous application to describe the structural transitions between B-DNA and two stretched conformations,  underwound S-DNA and overwound P-DNA.
The default control variables are tension $J$ and torque $\tau$.
All results are convertible into the functional relations directly probed by experiments \cite{ABLC98, LRS+99, SLCM01, BSG+03, CLH+96, CYL+04}.

At low torque, a gradual increase in tension is known to trigger a transition from B-DNA to S-DNA.
In Sec.~V of Ref.~\cite{mct1} we have already analyzed this transition in the absence of any torsional constraints.
At moderately low tension, a gradual increase in torque is known to  convert B-DNA into P-DNA.
At low tension this scenario is complicated by the formation of plectonemes, a phenomenon investigated in Sec.~\ref{sec:DNA-plect}.

\subsection{Nucleation and growth of conformations}\label{sec:nuc-gro-con}
In this application we employ two host/tag pairs of level-2 nested particles as introduced in Ref.~\cite{mct1} (Sec.~II.D and App. C). 
From the B-DNA reference state, segments of either S-DNA or P-DNA are nucleated by the activation of host particles and then grown by the activation of tag particles.
The combinatorial specifications of all four particle species are summarized in Table~\ref{tab:3}.

\begin{table}[htb]
\caption{Capacity constants $A_m$ and statistical interaction coefficients $g_{mm'}$ for two host/tag pairs of level-2 particles.}\label{tab:3}
\begin{tabular}{lc|c} \hline\hline \rule[-2mm]{0mm}{6mm}
~~ & $m$~~ & ~~$A_m$  \\ \hline \rule[-2mm]{0mm}{6mm}
S-host~~ & 1~~ & ~~$N-1$ \\ \rule[-2mm]{0mm}{5mm}
S-tag~~ & 2~~ & ~~$0$ \\ \rule[-2mm]{0mm}{5mm}
P-host~~ & 3~~ & ~~$N-1$ \\ \rule[-2mm]{0mm}{5mm}
P-tag~~ & 4~~ & ~~$0$ \\ \hline\hline 
\end{tabular} \hspace{5mm} 
\begin{tabular}{c|rrrr}\hline\hline  \rule[-2mm]{0mm}{6mm}
 $g_{mm'}$~ &  ~1  & ~~2 & 3 & ~~4 \\ \hline \rule[-2mm]{0mm}{6mm}
	1 & $2$ & $1$ & $1$ & $1$\\ \rule[-2mm]{0mm}{5mm}
	2 & $-1$ & $0$ & $0$ & $0$\\  \rule[-2mm]{0mm}{5mm}
	3 & $2$ & $1$ & $2$ & $1$\\ \rule[-2mm]{0mm}{5mm}
	4 & $0$ & $0$ & $-1$ & $0$\\  \hline\hline 
\end{tabular} 
\end{table}

The particle activation energies exhibit the standard dependence (\ref{eq:7}) on tension and torque,
\begin{subequations}\label{eq:14}
\begin{align} \label{eq:14a}
  \epsilon_{1} &=\gamma_\mathrm{S}- JL_\mathrm{S} -\tau \phi_\mathrm{S}+ c_\mathrm{S} , \\ \label{eq:14b}
 \epsilon_{2} &= \gamma_\mathrm{S} - JL_\mathrm{S} -\tau \phi_\mathrm{S}, 
 \\ \label{eq:14c}
  \epsilon_{3} &= \gamma_{P} - JL_\mathrm{P} -\tau \phi_\mathrm{P}+ c_\mathrm{P}, \\ \label{eq:14d}
  \epsilon_{4} &= \gamma_\mathrm{P} - JL_\mathrm{P} -\tau \phi_\mathrm{P},
\end{align}
\end{subequations}
here amended with constants  $c_\mathrm{S}$ or $c_\mathrm{P}$ in the two host species, by which the cooperativity of the conformational changes is being controlled.
All specifications used in this application are inferred from well established empirical data such as found in Refs.~\cite{ABLC98, LRS+99, SLCM01, CLH+96, CYL+04}:
\begin{subequations}\label{eq:21}
\begin{align}\label{eq:21a}
& L_\mathrm{S}=L_\mathrm{P}\doteq L_\mathrm{c}=0.24\mathrm{nm/bp}, 
\\ \label{eq:21b}
& \phi_\mathrm{S}=-0.42\mathrm{rad/bp},\quad 
\phi_\mathrm{P}=+1.5\mathrm{rad/bp}, \\ \label{eq:21c}
& \gamma_\mathrm{S}=16\mathrm{pNnm},\quad
\gamma_\mathrm{P}=60\mathrm{pNnm}.
\end{align}
\end{subequations}
The twist angles in this application are relative to the native chirality of B-DNA.

\subsection{Phase diagram}\label{sec:BSP-pha-dia}
\begin{figure}[htb]
\begin{center}
\includegraphics[width=41mm]{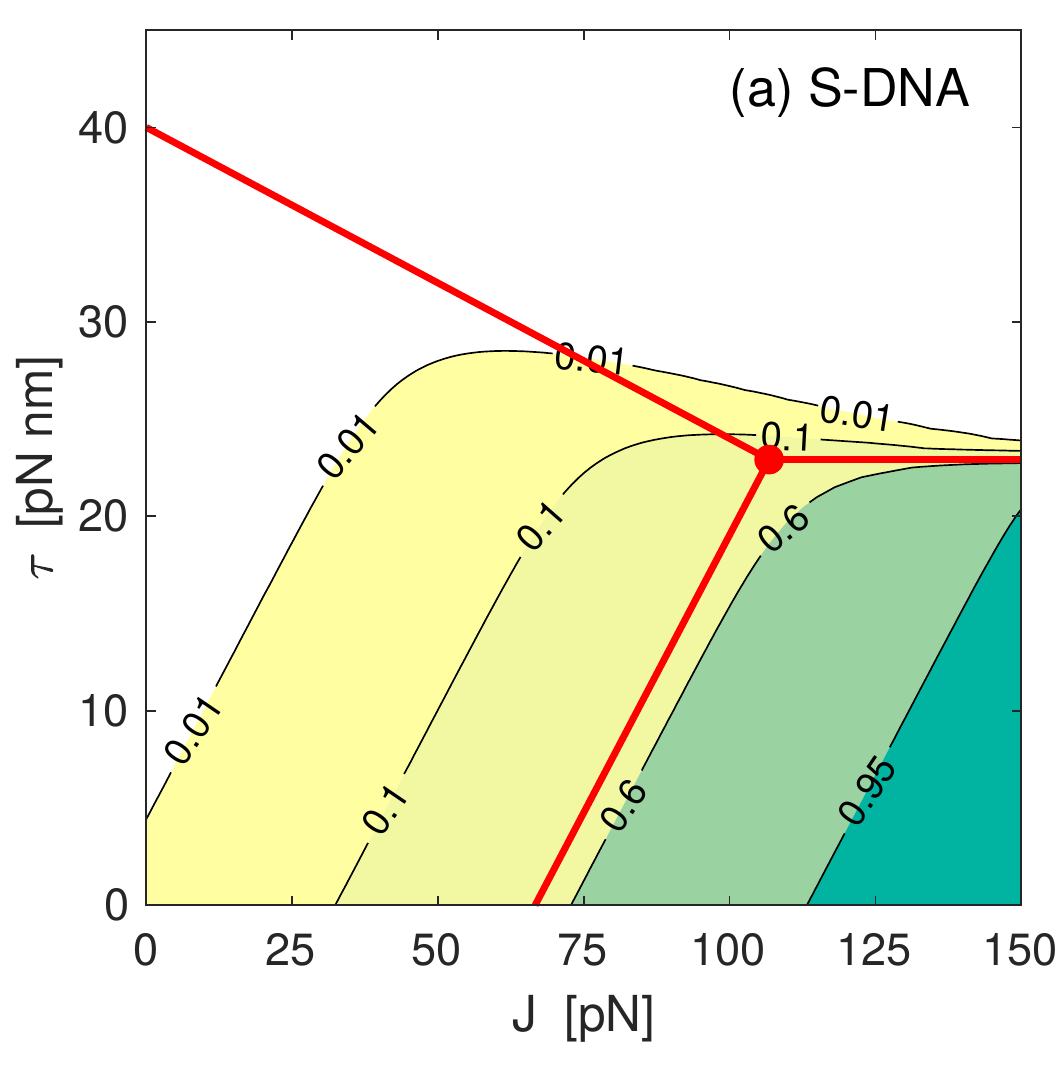}\hspace*{3mm}%
\includegraphics[width=41mm]{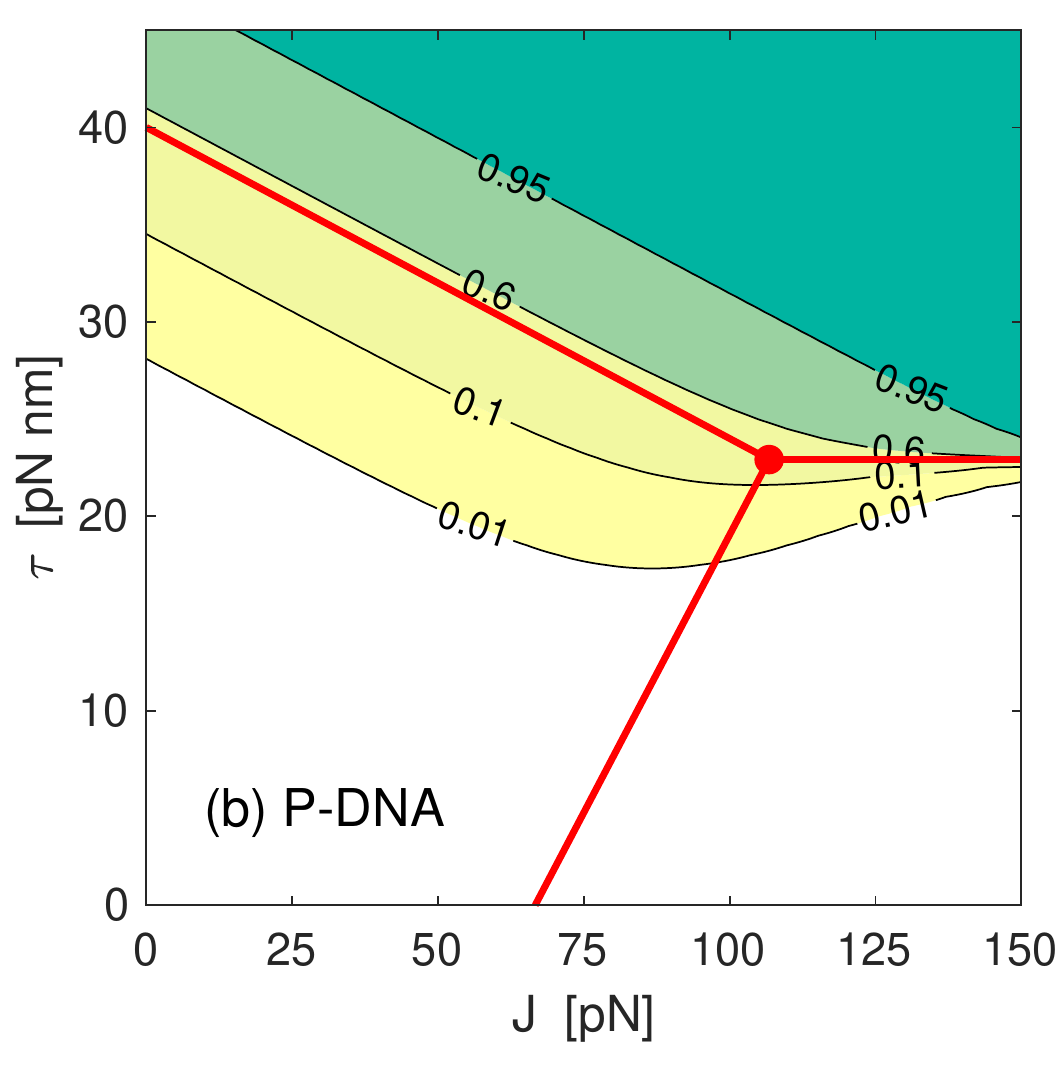}
\includegraphics[width=41mm]{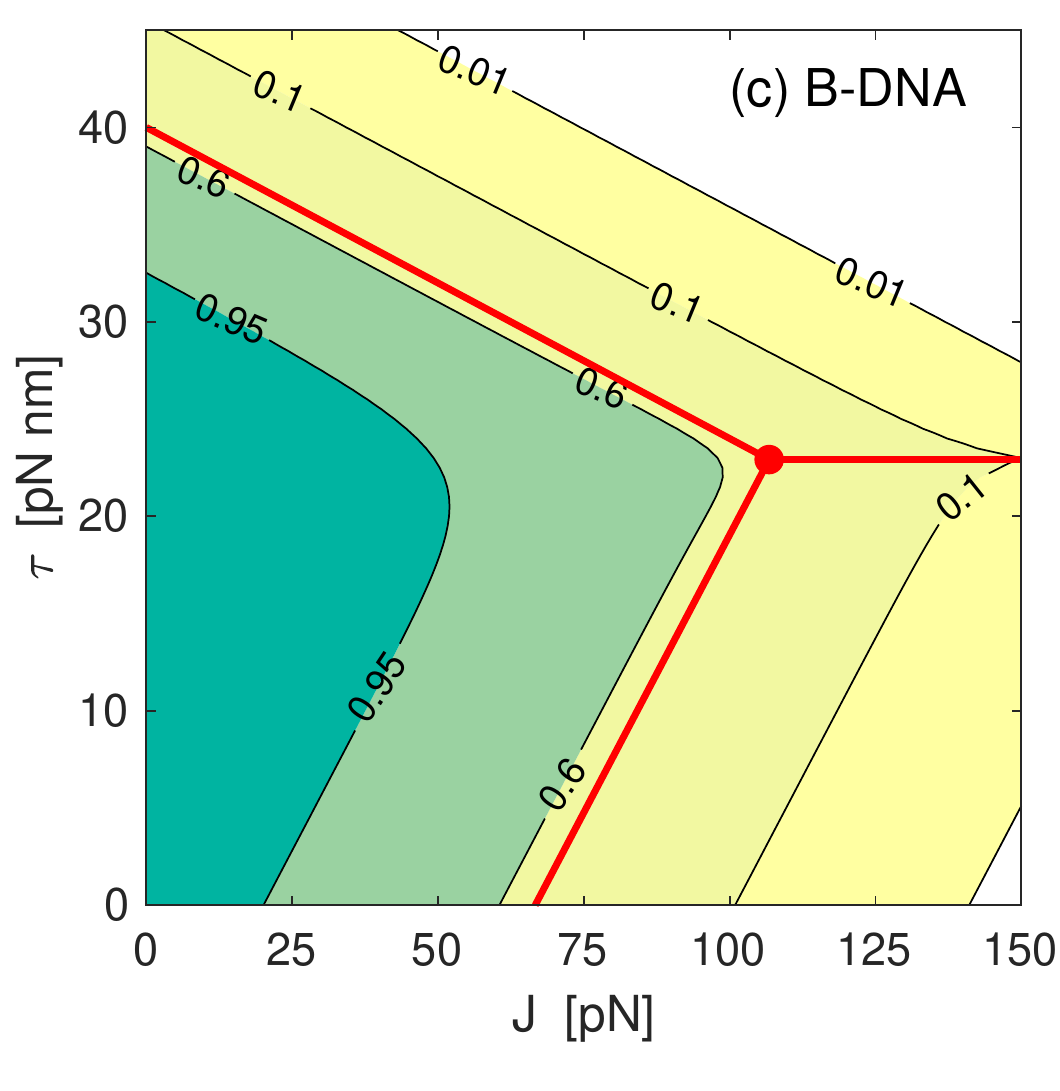}\hspace*{3mm}%
\includegraphics[width=41mm]{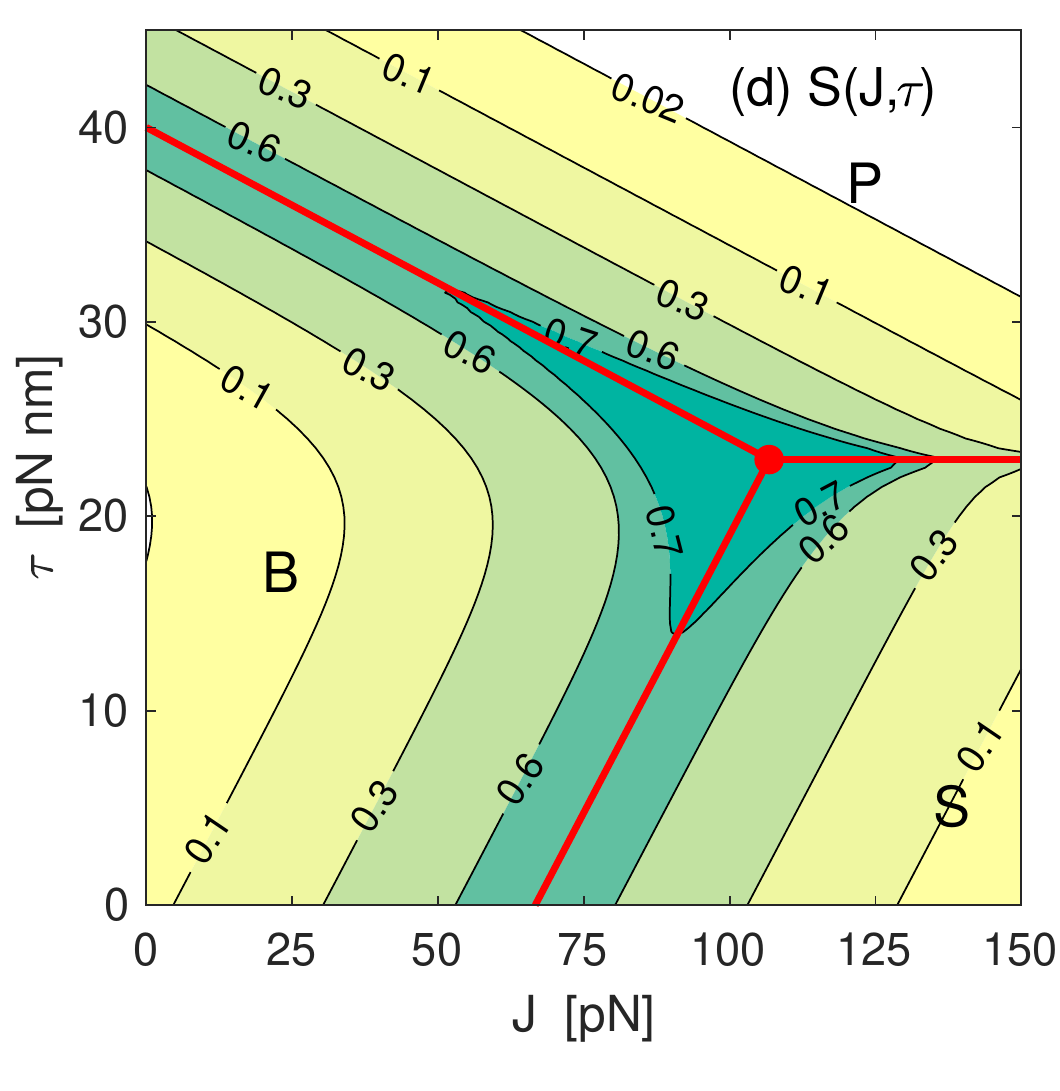}
\end{center}
\caption{Fraction of segments in the conformations of (a) S-DNA, (b) P-DNA,
 and (c) B-DNA.  (d) Scaled entropy.
 The parameter values $c_\mathrm{S}=c_\mathrm{P}=1$ pN nm indicate low
  cooperativity. 
  The (asymptotic) phase boundaries (\ref{eq:22}) are shown as straight lines, meeting in the triple point (\ref{eq:23}).}
		\label{fig:figure21}
	\end{figure}

\begin{figure}[!ht]
\begin{center}
\includegraphics[width=41mm]{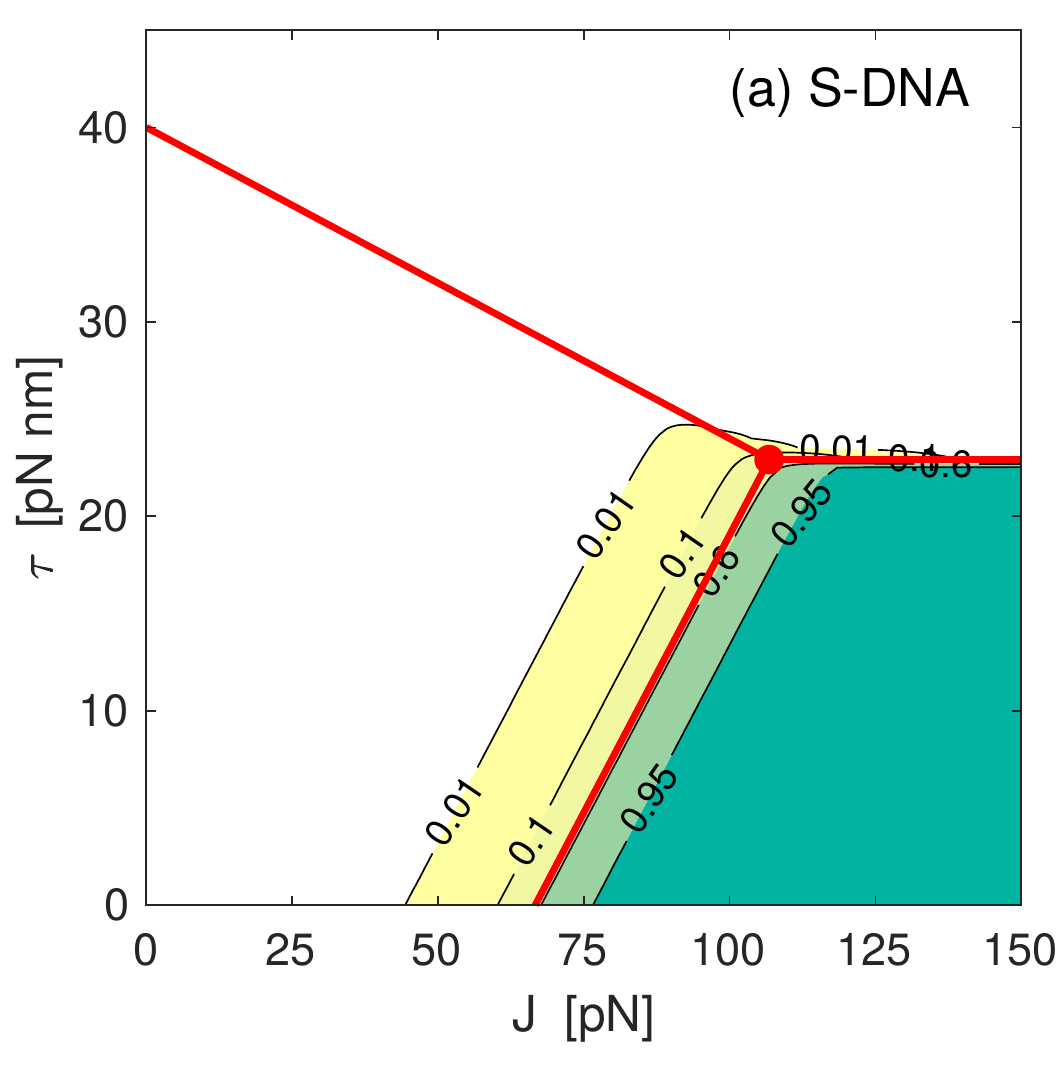}\hspace*{3mm}%
\includegraphics[width=41mm]{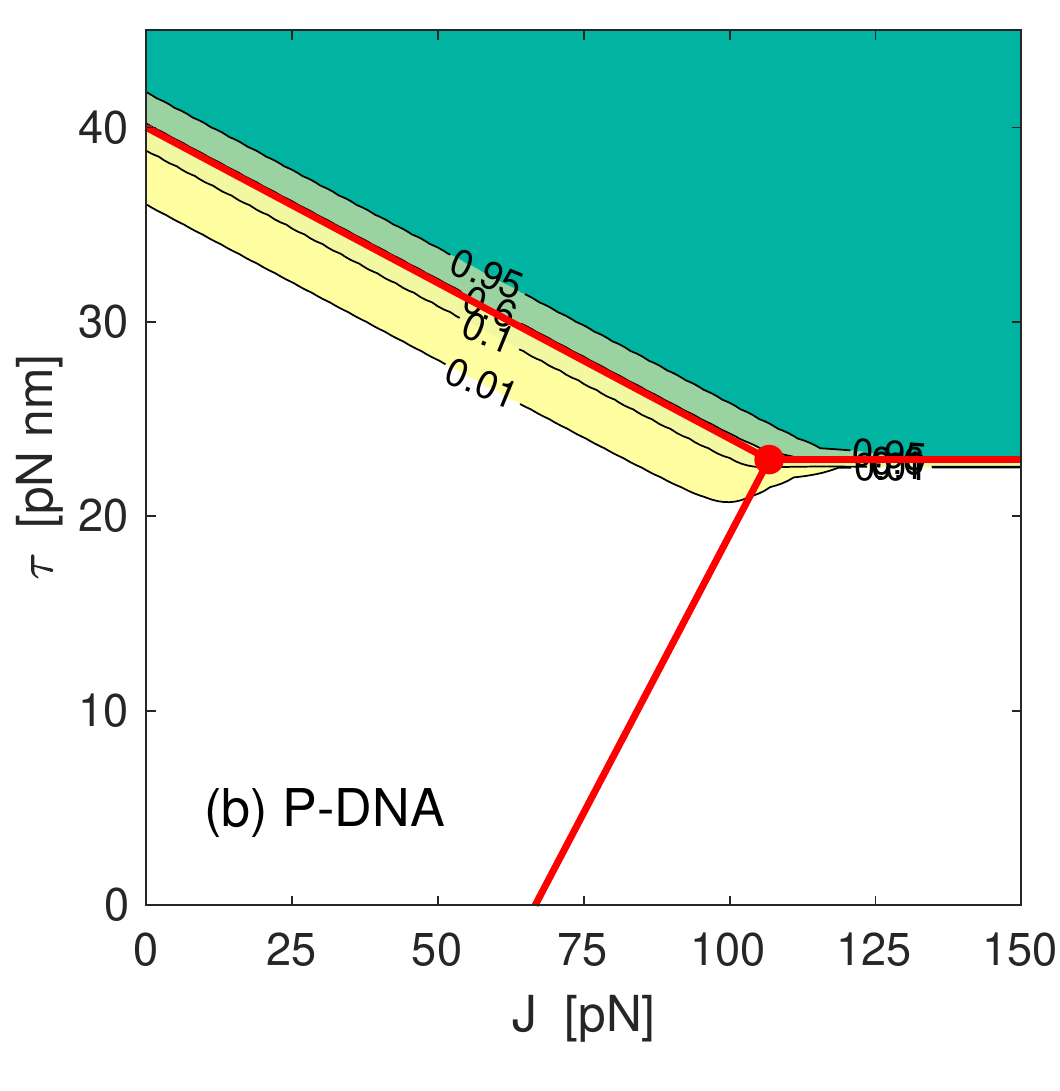}
\includegraphics[width=41mm]{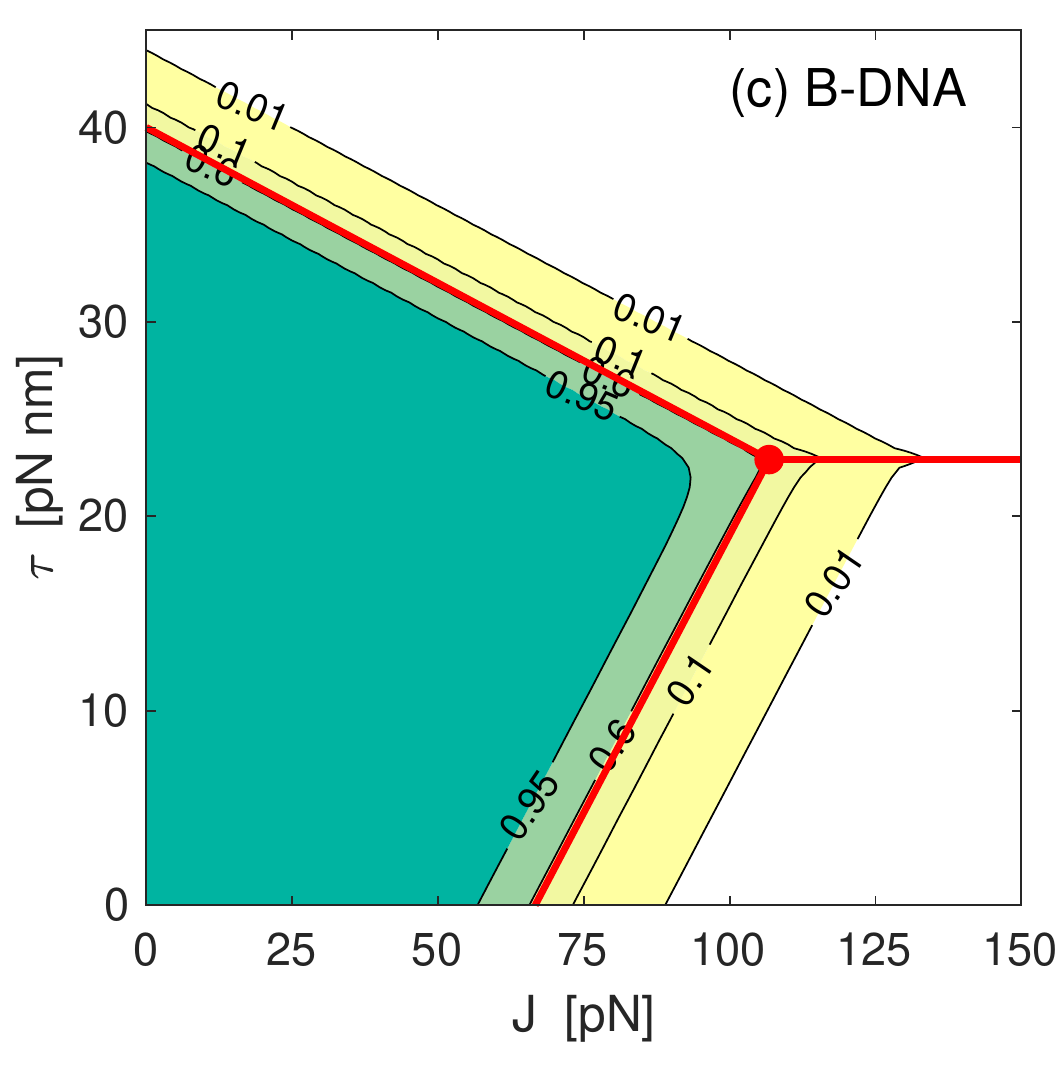}\hspace*{3mm}%
\includegraphics[width=41mm]{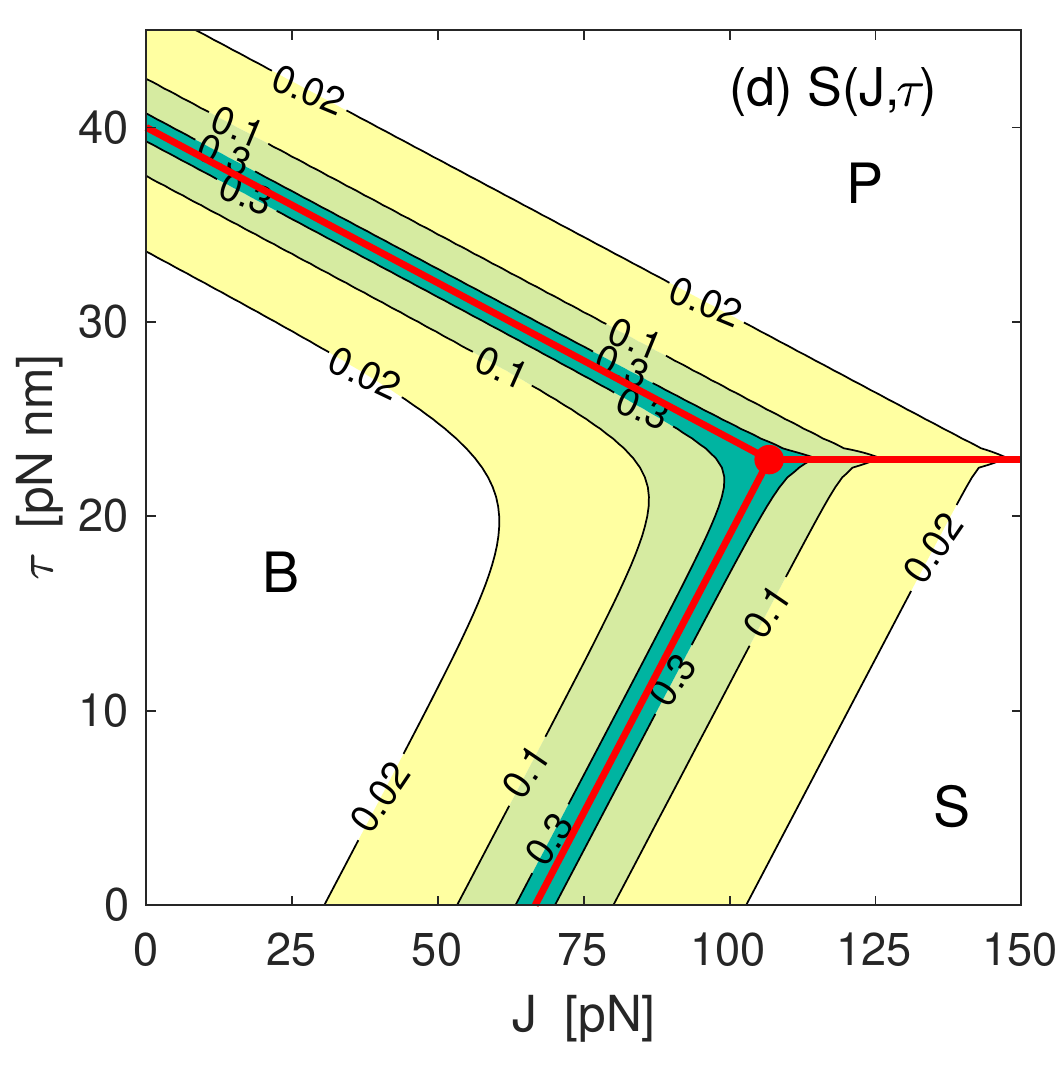}
\end{center}
\caption{Fraction of segments in the conformations of (a) S-DNA, (b) P-DNA,
  and (c) B-DNA.  (d) Scaled entropy.
  The parameter values $c_\mathrm{S}=16$ pN nm, $c_\mathrm{P}=15$ pN nm
  indicate high cooperativity.
  The (asymptotic) phase boundaries (\ref{eq:22}) are shown as straight lines, meeting in the triple point (\ref{eq:23}).}
\label{fig:figure22}
\end{figure}

The statistical mechanical analysis proceeds as in previous applications.
We calculate the particle population densities, $\bar{N}_m\doteq\lim_{N\to\infty}\langle N_m\rangle/N$, by solving Eqs.~(\ref{eq:5}) and (\ref{eq:6}) numerically and infer the scaled entropy, $\bar{S}(T,J,\tau)$, via Eqs.~(\ref{eq:17}). 
From the particle population densities $\bar{N}_m$, $m=1,\ldots,4$, we infer the fraction of ds-DNA in each conformation as follows:
\begin{align}\label{eq:15} 
&F_\mathrm{S}=\bar{N}_1+\bar{N}_2,\quad 
F_\mathrm{P}=\bar{N}_3+\bar{N}_4, \nonumber \\
&F_\mathrm{B}=1-F_\mathrm{S}-F_\mathrm{P.}
\end{align}
Contour plots of these fractions are shown in Fig.~\ref{fig:figure21}(a)-(c) for low cooperativity and in Fig.~\ref{fig:figure22}(a)-(c) for high cooperativity.
Three regions of predominant conformation are clearly identifiable. 
The crossover regions are broad at low cooperativity.
At high cooperativity, they look more like three phase boundaries ending in one vertex.

As a consistency check we have added panel (d) to Figs.~\ref{fig:figure21} and \ref{fig:figure22}, which shows a contour plot of the scaled entropy, $\bar{S}/k_\mathrm{B}$.
The level of disorder is significantly higher in regions where two conformations mix and even higher where all three conformations are present.
The regions of mixed conformation become narrow when cooperativity is high.
Here the lines of enhanced entropy turn into accurate markers of the three emergent phase boundaries already identified.
At the P-S border, which is realized at high tension, little entropy is produced. 
However, this does not reduce the accuracy of the entropy as a phase-boundary locator.

The B-S phase boundary at zero torque and the B-P phase boundary at very low tension, both well-defined at high cooperativity, are consistent with experimental benchmarks \cite{ABLC98, LRS+99, SLCM01, CLH+96, CYL+04}.
The former is also in quantitative agreement with one relevant landmark in a different comparison of experimental data with results from our methodology, namely in the context of torsion-free stretching \cite{SACB00, mct1}.

The positive slope of the B-S phase boundary is attributable to the fact that S-DNA is underwound.
Stretching at higher torque delays the B-S transition.
The negative slope of the B-P phase boundary is explained by the fact that P-DNA is overwound.
Stretching at higher torque enhances the B-P transition.
The emergent vertex at $\tau\simeq 23$ pN nm and $J\simeq 110$ pN, which is identifiable in 
Fig.~\ref{fig:figure22}(d) with fair precision, is well established experimentally in the form of a secondary force-extension plateau found in torsionally constrained DNA \cite{SAB+96, Wang97}.

\subsection{Transition asymptotics}\label{sec:tra-asy}
Here present the analytical form for the emergent phase boundaries, derived from the partition function (\ref{eq:4}), which essentially informs all results in every application.
This extraction is somewhat technical, carried out for room temperature in a high-cooperativity limit.
The analysis yields three phase boundaries in the form of straight line segments in the $(J,\tau)$-plane meeting in one vertex:
\begin{subequations}\label{eq:22}
\begin{equation}\label{eq:22a} 
 \tau_\mathrm{BS}(J)=\frac{JL_\mathrm{c}-\gamma_\mathrm{S}}{|\phi_\mathrm{S}|}
 \quad :~ \frac{\gamma_\mathrm{S}}{L_\mathrm{c}}\leq J\leq J_\mathrm{T},
 \end{equation} 
 \begin{equation}\label{eq:22b} 
 \tau_\mathrm{BP}(J)=\frac{\gamma_\mathrm{P}-JL_\mathrm{c}}{\phi_\mathrm{P}}
 \quad :~ 0\leq J\leq J_\mathrm{T},
 \end{equation}
 \begin{equation}\label{eq:22c} 
 \tau_\mathrm{SP}(J)=\tau_\mathrm{T}
 \quad :~ J\geq J_\mathrm{T},
 \end{equation}
 \end{subequations}
   \begin{subequations}\label{eq:23} 
  \begin{equation}\label{eq:23a} 
 J_\mathrm{T}=\frac{\gamma_\mathrm{S}\phi_\mathrm{P}
 -\gamma_\mathrm{P}\phi_\mathrm{S}}
 {L_\mathrm{c}(\phi_\mathrm{P}-\phi_\mathrm{S})}= 106.8\mathrm{pN},
 \end{equation}
  \begin{equation}\label{eq:23b} 
 \tau_\mathrm{T}=\frac{\gamma_\mathrm{P}-\gamma_\mathrm{S}}
 {\phi_\mathrm{P}-\phi_\mathrm{S}}= 22.9\mathrm{pNnm}. 
 \end{equation}
 \end{subequations}
These phase boundaries, in accord with experimental data, are included in Figs.~\ref{fig:figure21} and \ref{fig:figure22}.

The conformational fractions are governed, for the most part, by the activation energies of the S-tags and P-tags of Table~\ref{tab:3}.
This is evident in the following leading-order asymptotic expression.
In the S-phase, where $\epsilon_2<0$ and $\epsilon_4>0$, we have
\begin{subequations}\label{eq:35}
\begin{align}\label{eq:35a}
F_\mathrm{S} &\simeq\frac{1}{2}\left( 
        1- \frac{\sinh(\beta \epsilon_{2}/2)}%
        {\sqrt{\sinh^{2}(\beta \epsilon_{2}/2)+e^{-\beta c_\mathrm{S}}}}
      \right),  \\ \label{eq:35b}
F_\mathrm{P} &=\mathrm{O}\big(e^{-\beta \epsilon_{3}}\big), 
\\ \label{eq:35c}
F_\mathrm{B} &\simeq \frac{1}{2}\left( 
        1+ \frac{\sinh(\beta \epsilon_{2}/2)}%
        {\sqrt{\sinh^{2}(\beta \epsilon_{2}/2)+e^{-\beta c_\mathrm{S}}}}
      \right). 
\end{align}
\end{subequations}
In the P-phase, we have $\epsilon_2>0$, $\epsilon_4<0$, yielding
\begin{subequations}\label{eq:36}
\begin{align}\label{eq:36a}
F_\mathrm{S} &=\mathrm{O}\big(e^{-\beta \epsilon_{1}}\big) 
\\ \label{eq:36b}
F_\mathrm{P} &\simeq \frac{1}{2}\left( 
        1- \frac{\sinh(\beta \epsilon_{4}/2)}%
        {\sqrt{\sinh^{2}(\beta \epsilon_{4}/2)+e^{-\beta c_\mathrm{P}}}}
      \right), \\ \label{eq:36c}
F_\mathrm{B} &\simeq \frac{1}{2}\left( 
        1+ \frac{\sinh(\beta \epsilon_{4}/2)}%
        {\sqrt{\sinh^{2}(\beta \epsilon_{4}/2)+e^{-\beta c_\mathrm{P}}}}
      \right). 
\end{align}
\end{subequations}
Positive activation energies, $\epsilon_2>0$ and $\epsilon_4>0$, suppress S-tags and P-tags in the B-phase:
\begin{subequations}\label{eq:37}
\begin{align}\label{eq:37a}
F_\mathrm{S} &\simeq e^{-\beta \epsilon_{2}}e^{-\beta c_\mathrm{S}},
\\ \label{eq:37b}
F_\mathrm{P} &\simeq e^{-\beta \epsilon_{4}}e^{-\beta c_\mathrm{P}}, 
\\ \label{eq:37c}
F_\mathrm{B} &\simeq 1-e^{-\beta \epsilon_{2}}e^{-\beta c_\mathrm{S}}
-e^{-\beta \epsilon_{4}}e^{-\beta c_\mathrm{P}}. 
\end{align}
\end{subequations}
The results (\ref{eq:35b}) and (\ref{eq:36a}) mean that those fractions are exponentially suppressed throughout that particular phase.
These functional forms in combination with specifications (\ref{eq:21}) are testable predictions.

%
\section{Supercoils}\label{sec:sup-coi}
%
We begin our analysis of supercoils by returning to the idealized molecular chain with a ladder-like reference state introduced in Sec.~\ref{sec:twis-cont}.
We now consider the possibility of two different responses to torque: the twisting of bonds or monomers and the formation of supercoils.
We distinguish two scenarios, one pertaining to high tension and the other to low tension.

The first scenario (Sec.~\ref{sec:sup-coi-ht}) describes a system with low torsional stiffness.
When a torque of increasing strength is applied under significant (constant) tension, the system first responds via twist contraction as described in Sec.~\ref{sec:twis-cont}.
When the twist response approaches saturation, the system is in need of a new way to respond to yet stronger torque.
That mode involves the formation of supercoils, which in this case are loops of highly twisted chain (as in a rope).

The second scenario (Sec.~\ref{sec:sup-coi-lt}) describes a system with high torsional stiffness.
At low (constant) tension, the primary response to an applied torque of increasing strength is the formation of supercoils, which in this case are loops of largely untwisted chain (as in a garden hose).
When we increase the tension under torsional constraint, supercoiled chain is gradually converted into twisted chain. 
With some modifications, either scenario is adaptable to DNA applications.
One realization of a buckling transition is demonstrated in Sec.~\ref{sec:DNA-plect}. 

\begin{figure}[htb]
  \begin{center}
\includegraphics[width=80mm]{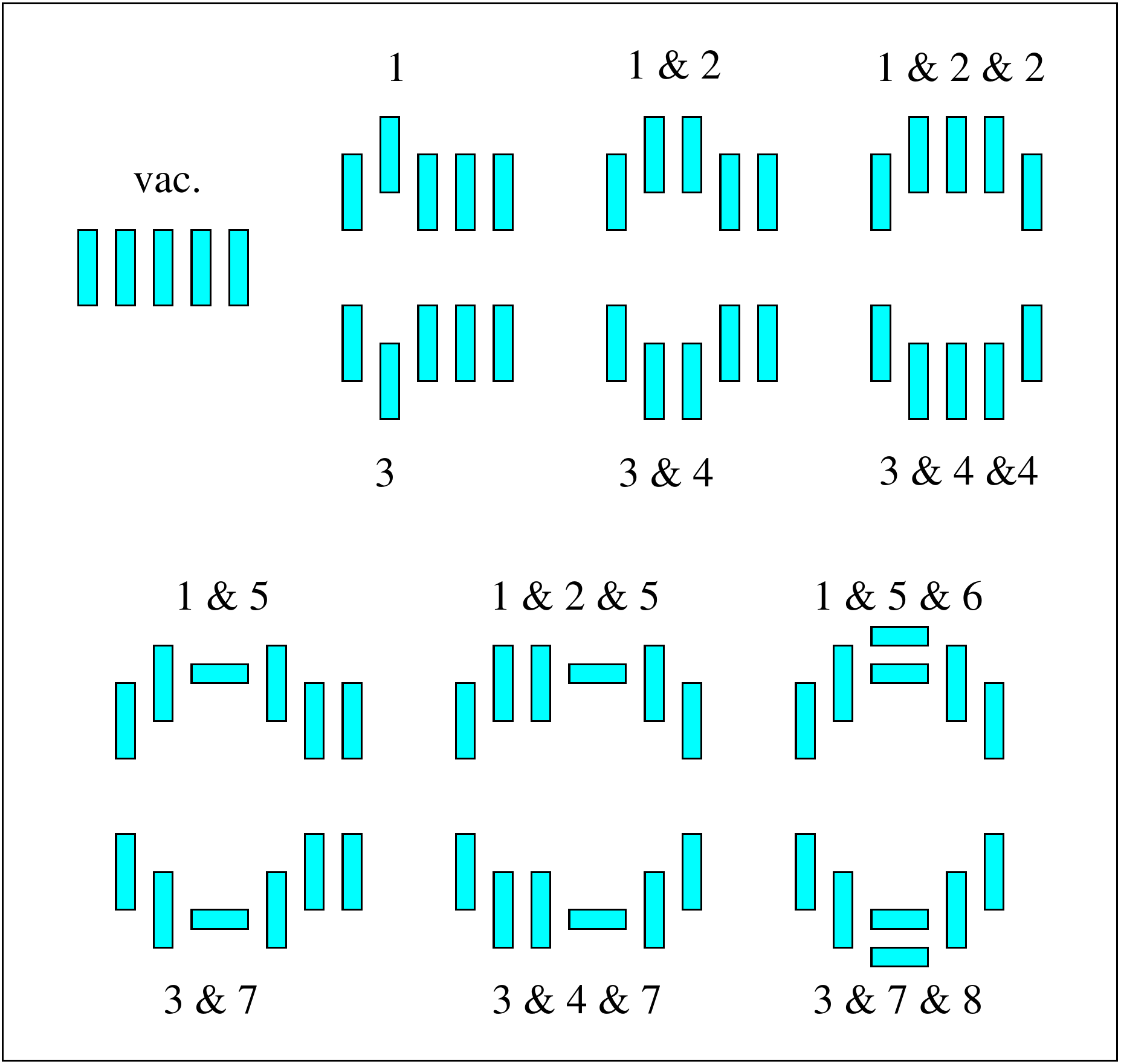}
\end{center}
\caption{Four species of level-2 twist particles (hosts 1, 3 and tags 2, 4) and four species of supercoil particles (hybrids 5, 7 and tags 6, 8) constituting two nested structures (one for each sense of chirality). Positive (negative) torque lowers the activation energies of particles 1, 2, 5, 6 (3, 4, 7, 8).}
  \label{fig:figure11}
\end{figure}

The modeling of supercoil conformations employs level-2 nested particles as introduced in  App.~C of \cite{mct1}.
Chain segments of left-handed or right-handed twist are activated by twist particles and segments of left-handed or right-handed writhe by supercoil particles.
We use four species (a host, a hybrid, and two tags) for each sense of chirality.
Symbolic representations of the eight species of particles are shown in Fig.~\ref{fig:figure11}.

Hosts 1 and 3 nucleate segments of twisted chain with opposite chirality out of the untwisted ladder reference state.
Tags 2 and 4 contribute to the growth of twisted-chain segments.
Hybrids 5 and 7 nucleate segments of supercoil.
The growth of supercoil segments is governed by tags 6 and 8.
In the first scenario (Sec.~\ref{sec:sup-coi-ht}), the supercoil segments grow out of twisted chain (e.g. by tags 6 replacing tags 2), whereas in the second scenario (Sec.~\ref{sec:sup-coi-lt}) they grow out of untwisted chain (e.g. by tags 6 replacing elements of reference state).
The level-2 particles used here offer two key advantages over the level-1 particles used in Sec.~\ref{sec:twis-cont}. They provide a parameter which controls the cooperativity of twisted chain segments and they facilitate the model extension to include supercoils.

The combinatorial analysis as previously described in different contexts \cite{LVP+08,copic,picnnn,pichs,cohetra} yields the specifications compiled in Table~\ref{tab:2}.
The dependence on tension and torque of the particle activation energies has the standard form (\ref{eq:7}), here again rewritten with a change of sign for convenience:
\begin{equation}\label{eq:18} 
\epsilon_m=\gamma_m+JL_m-\tau\phi_m.
\end{equation}

\begin{table}[htb]
  \caption{Statistical interaction coeffficients $g_{mm'}$ of the four species of twist (\textsf{tw}) particles and the four species of supercoil (\textsf{sc}) particles. The capacity constants are $A_1=A_3=N-2$ and $A_m=0$, $m=2, 4, 5, 6, 7, 8$.}\label{tab:2} 
  \begin{tabular}{lc|rrrrrrrr}\hline\hline  \rule[-2mm]{0mm}{6mm}
$g_{mm'}$ & $m\big\backslash m'$ & $1$ & $~~2$ & $3$ & $~~4$  & $~~5$ & $~~6$ & $~~7$ & $~~8$\\ \hline \rule[-2mm]{0mm}{6mm}
\textsf{tw} host & $1$ & $2$ & $1$ & $2$ & $1$ & $2$ & $1$ & $2$ & $1$\\ \rule[-2mm]{0mm}{5mm}
\textsf{tw} tag & $2$ & $-1$ & $0$ & $0$ & $0$ & $-1$ & $0$ & $0$ & $0$\\  \rule[-2mm]{0mm}{5mm}
\textsf{tw} host & $3$ & $1$ & $1$ & $2$ & $1$ & $2$ & $1$ & $2$ & $1$\\ \rule[-2mm]{0mm}{5mm}
\textsf{tw} tag & $4$ & $0$ & $0$ & $-1$ & $0$ & $0$ & $0$ & $-1$ & $0$\\  \rule[-2mm]{0mm}{5mm}
\textsf{sc} hybrid & $5$ & $-1$ & $0$ & $0$ & $0$ & $0$ & $0$ & $0$ & $0$\\ \rule[-2mm]{0mm}{5mm}
\textsf{sc} tag & $6$ & $0$ & $0$ & $0$ & $0$ & $-1$ & $0$ & $0$ & $0$\\  \rule[-2mm]{0mm}{5mm}
\textsf{sc} hybrid & $7$ & $0$ & $0$ & $-1$ & $0$ & $0$ & $0$ & $0$ & $0$\\ \rule[-2mm]{0mm}{5mm}
\textsf{sc} tag & $8$ & $0$ & $0$ & $0$ & $0$ & $0$ & $0$ & $-1$ & $0$\\ \hline\hline 
\end{tabular} 
\end{table} 

The two kinds of nucleation and growth processes are controlled by the parameters $\gamma_m, L_m, \phi_m$.
We set 
\begin{align}\label{eq:25} 
& \gamma_m>0,\quad L_m>0\quad:~ m=1,\ldots,8, \nonumber \\
& \phi_m>0\quad :~m=1,2,5,6, \\
& \phi_m<0\quad :~m=3,4,7,8. \nonumber 
\end{align}
Symmetry considerations require the following relations between parameters: 
\begin{align}\label{eq:19} 
& \gamma_1=\gamma_3,\quad \gamma_2=\gamma_4,\quad \gamma_5=\gamma_7,\quad \gamma_6=\gamma_8, \nonumber \\
& L_1=L_3,\quad L_2=L_4,\quad L_5=L_7,\quad L_6=L_8,  \\
& \phi_1=-\phi_3,\quad \phi_2=-\phi_4,\quad \phi_5=-\phi_7,\quad \phi_6=-\phi_8. \nonumber
\end{align}
In applications to chains with intrinsic chirality these relations need to be modified as will be explored in a separate study \cite{mct3}.

%
\section{Supercoils at high tension}\label{sec:sup-coi-ht}
%
This first scenario is designed (by parameter setting) to produce supercoil segments made of highly twisted chain. 
Under increasing torque, supercoil particles crowd out twist particles. 
The former thus incorporate two kinds of contraction lengths and two kinds of angles,
namely twist and writhe.
This is accommodated by specifications which satisfy the inequalities,
\begin{subequations}\label{eq:20} 
\begin{align}\label{eq:20a} 
& L_5>L_1,\quad L_7>L_3,\quad L_6>L_2,\quad L_8>L_4, \\ \label{eq:20b} 
& \phi_5>\phi_1,\quad |\phi_7|>|\phi_3|,\quad \phi_6>\phi_2,\quad |\phi_8|>|\phi_4|.
\end{align}
The additional inequalities,
\begin{equation}\label{eq:20c} 
\gamma_5>\gamma_1,\quad \gamma_7>\gamma_2,\quad \gamma_6>\gamma_2,\quad \gamma_8>\gamma_4,
\end{equation}
\end{subequations}
facilitate the empirical requirement that supercoiling is a phenomenon realized in a highly twisted chain.

\subsection{Low torque: twisted chain}\label{sec:low-tor-reg}
With the parameter setting satisfying the constraints (\ref{eq:20}) there exists a regime of low torque where only twist particles are activated in significant numbers.
Here supercoil particles have high activation energies and are effectively frozen out.

In the statistical mechanical analysis, the limit of infinitely high activation energies for supercoil particles implies that
\begin{equation}\label{eq:26} 
w_m\to\infty \quad :~ m=5,\ldots,8,
\end{equation}
which, in turn, has the consequence that the linear Eqs.~(\ref{eq:6}) yield vanishing population densities for supercoil particles:
\begin{equation}\label{eq:27}
\bar{N}_m\doteq\frac{\langle N_m\rangle}{N}\to0 \quad :~ m=5,\ldots,8.
\end{equation}
The polynomial Eqs.~(\ref{eq:5}) for the remaining variables, $w_m$, $m=1,\ldots,4$, are then of lower order, cubic in this case.
The (unique) physical solution determines the population densities of twist particles via Eqs.~(\ref{eq:6}) for $m=1,\ldots,4$ and the Gibbs free energy $\bar{G}(T,J,\tau)$ via (\ref{eq:8}).
The contraction length $\bar{L}$, the entropy $\bar{S}$, and the twist angle $\bar{\phi}$ are calculated from from $\bar{G}(T,J,\tau)$ via derivatives as in (\ref{eq:9}) or via population densities via (\ref{eq:17}) and (\ref{eq:16}) properly rescaled.

\begin{figure}[t]
  \begin{center}
\includegraphics[width=40mm]{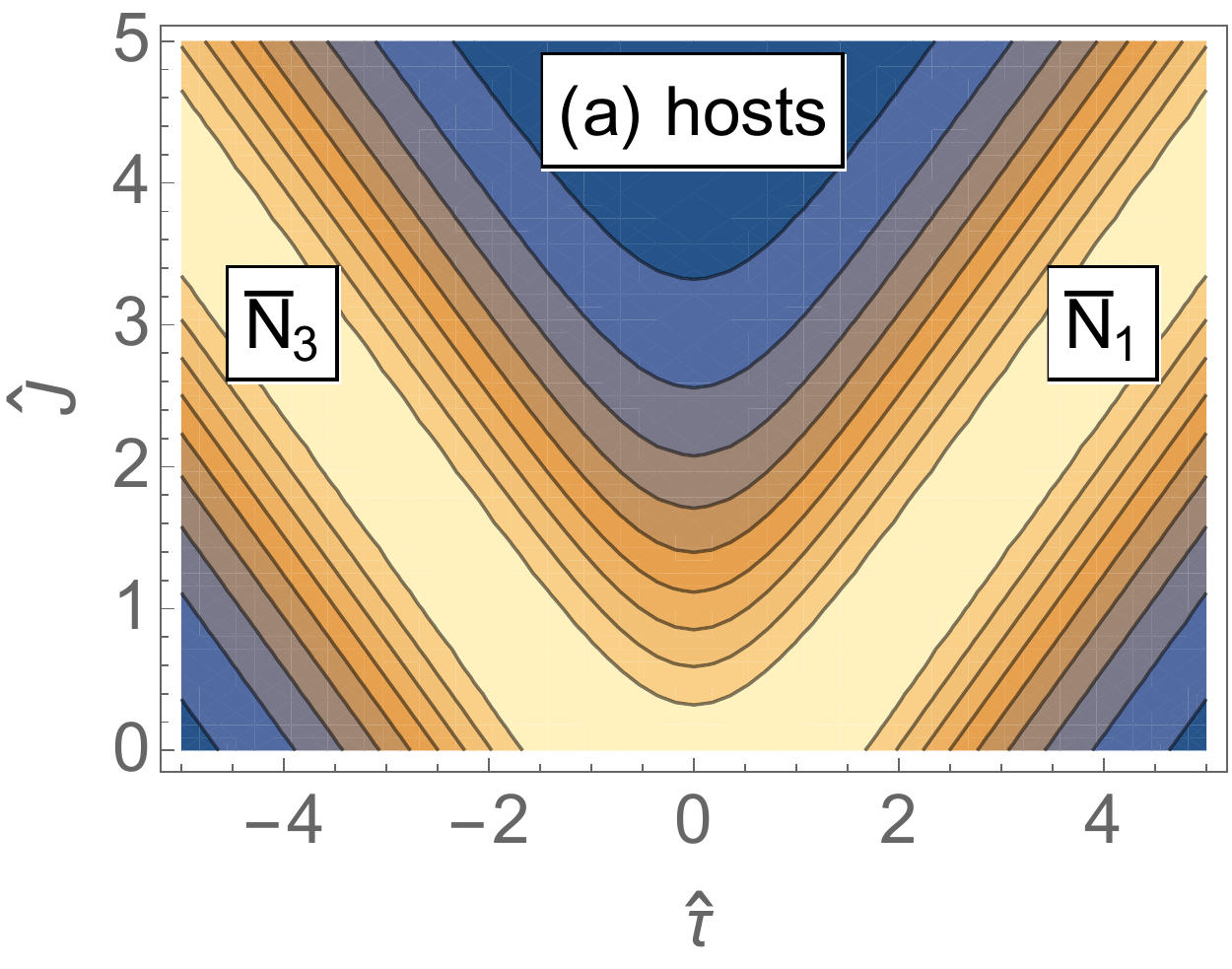}\hspace*{3mm}\includegraphics[width=40mm]{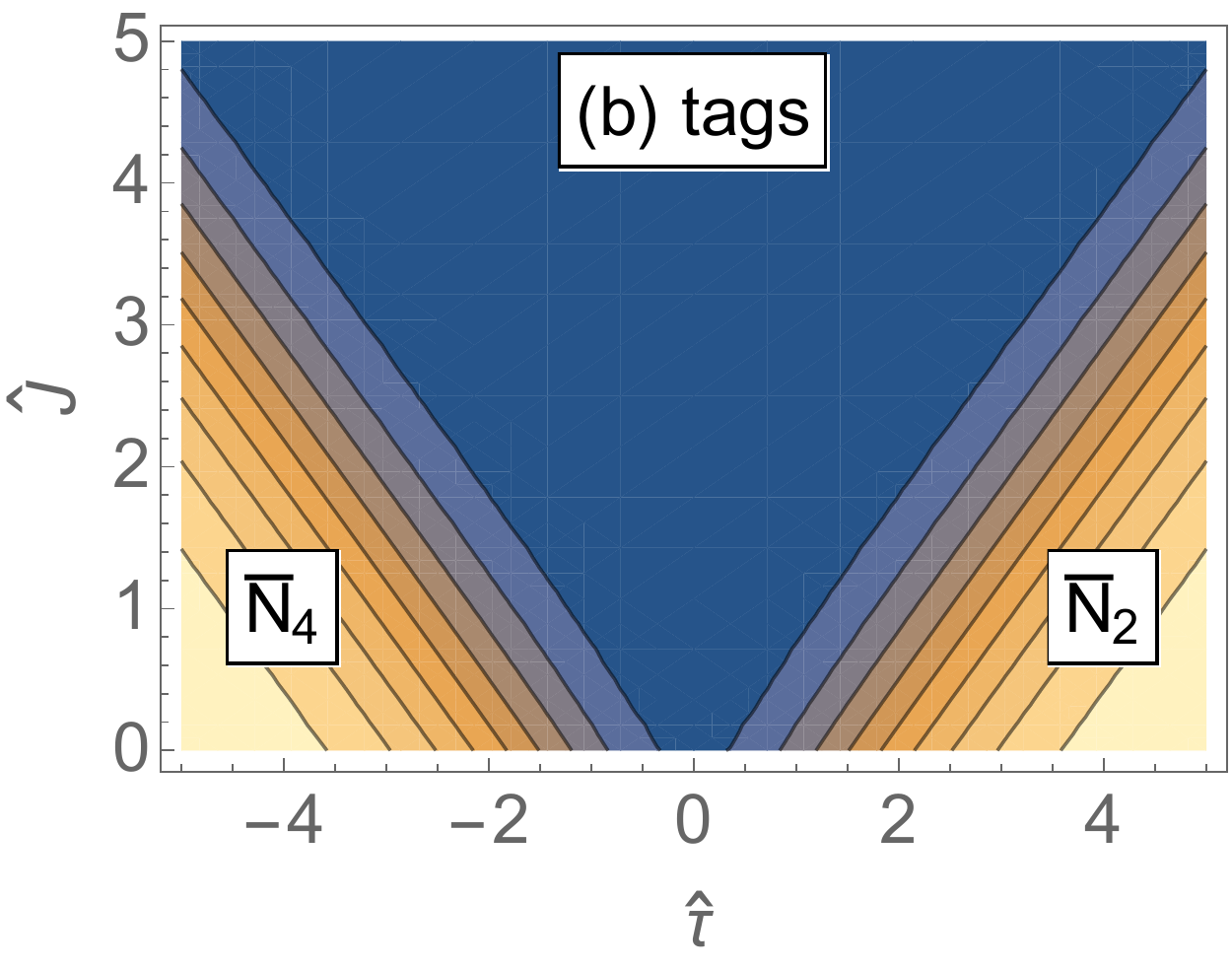}
\includegraphics[width=40mm]{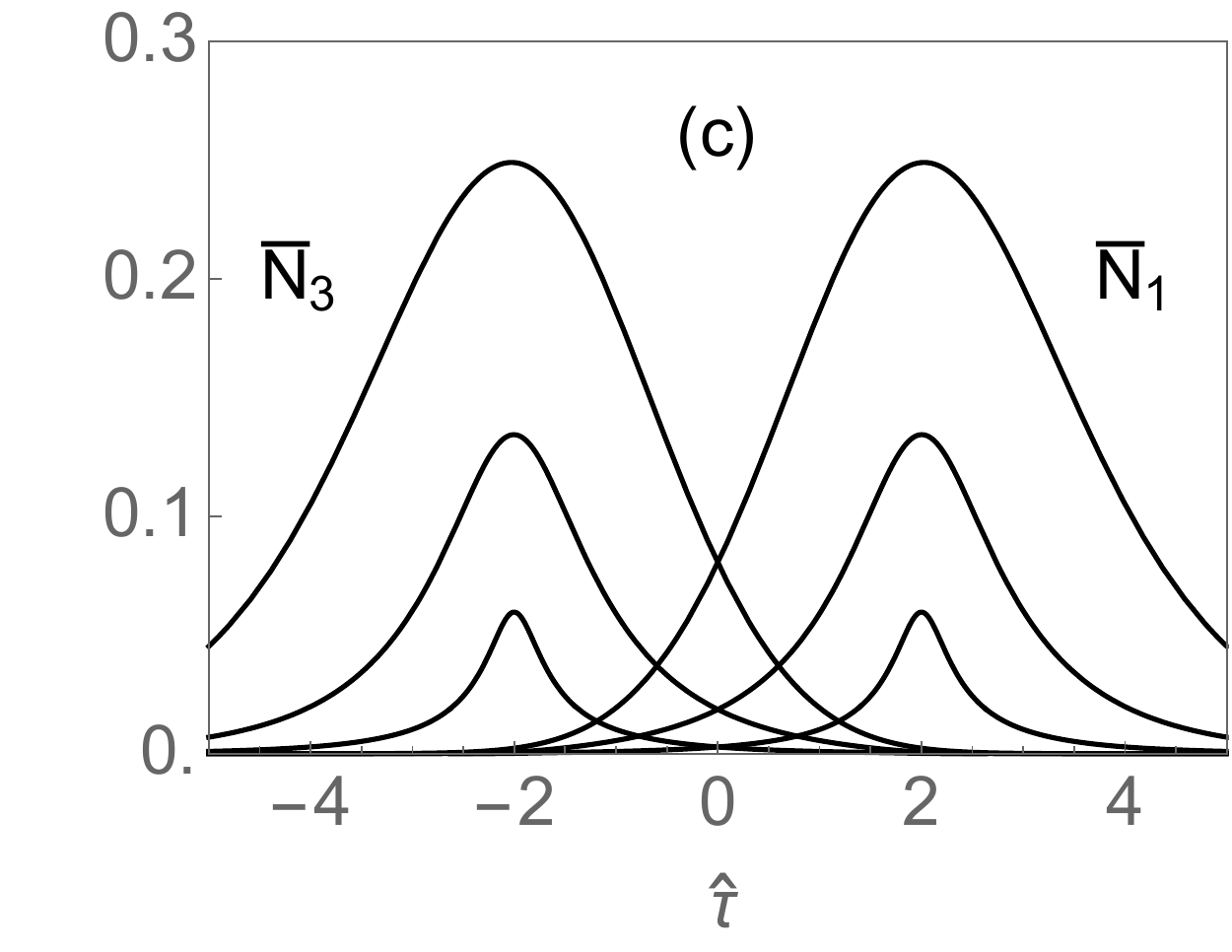}\hspace*{3mm}\includegraphics[width=40mm]{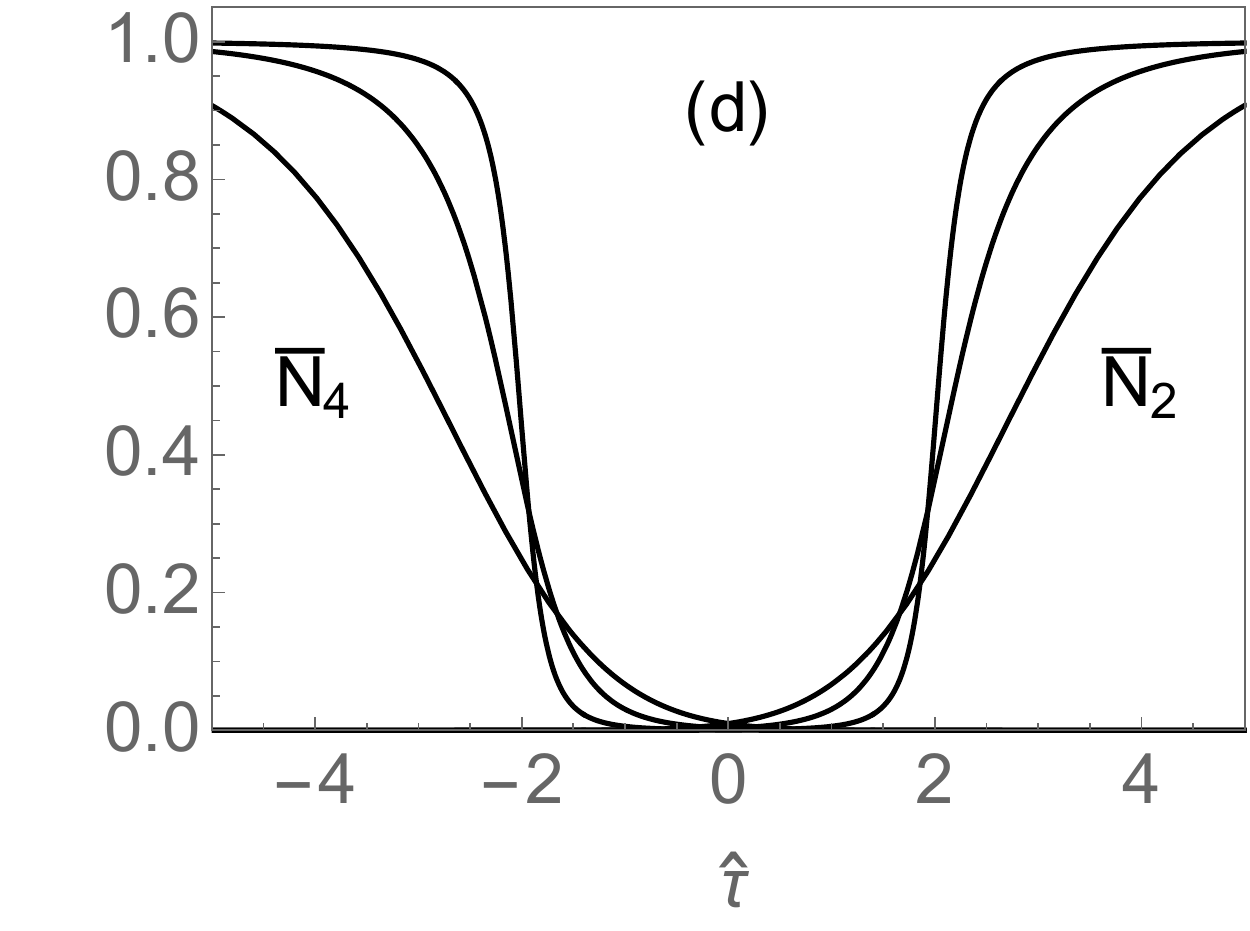}
\includegraphics[width=40mm]{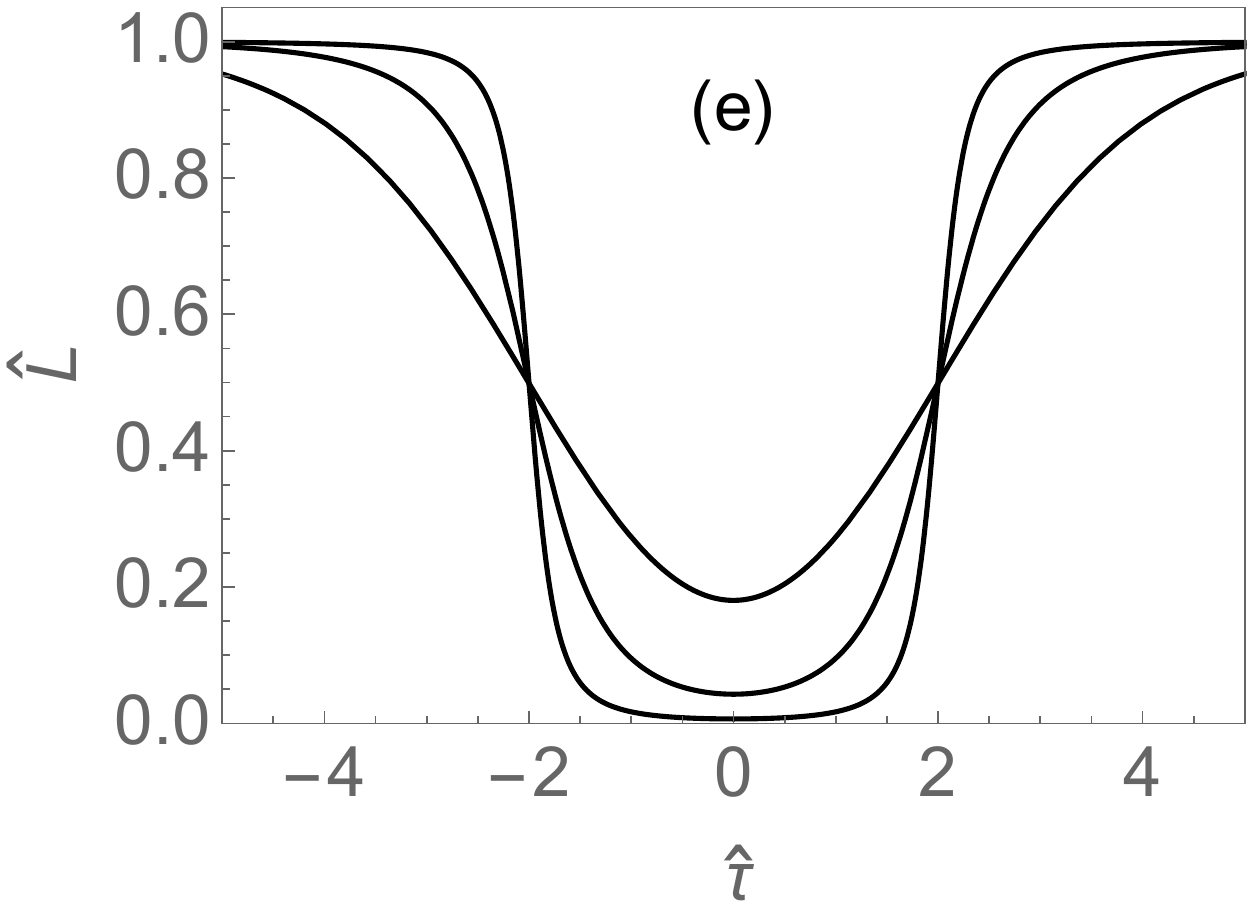}\hspace*{3mm}\includegraphics[width=40mm]{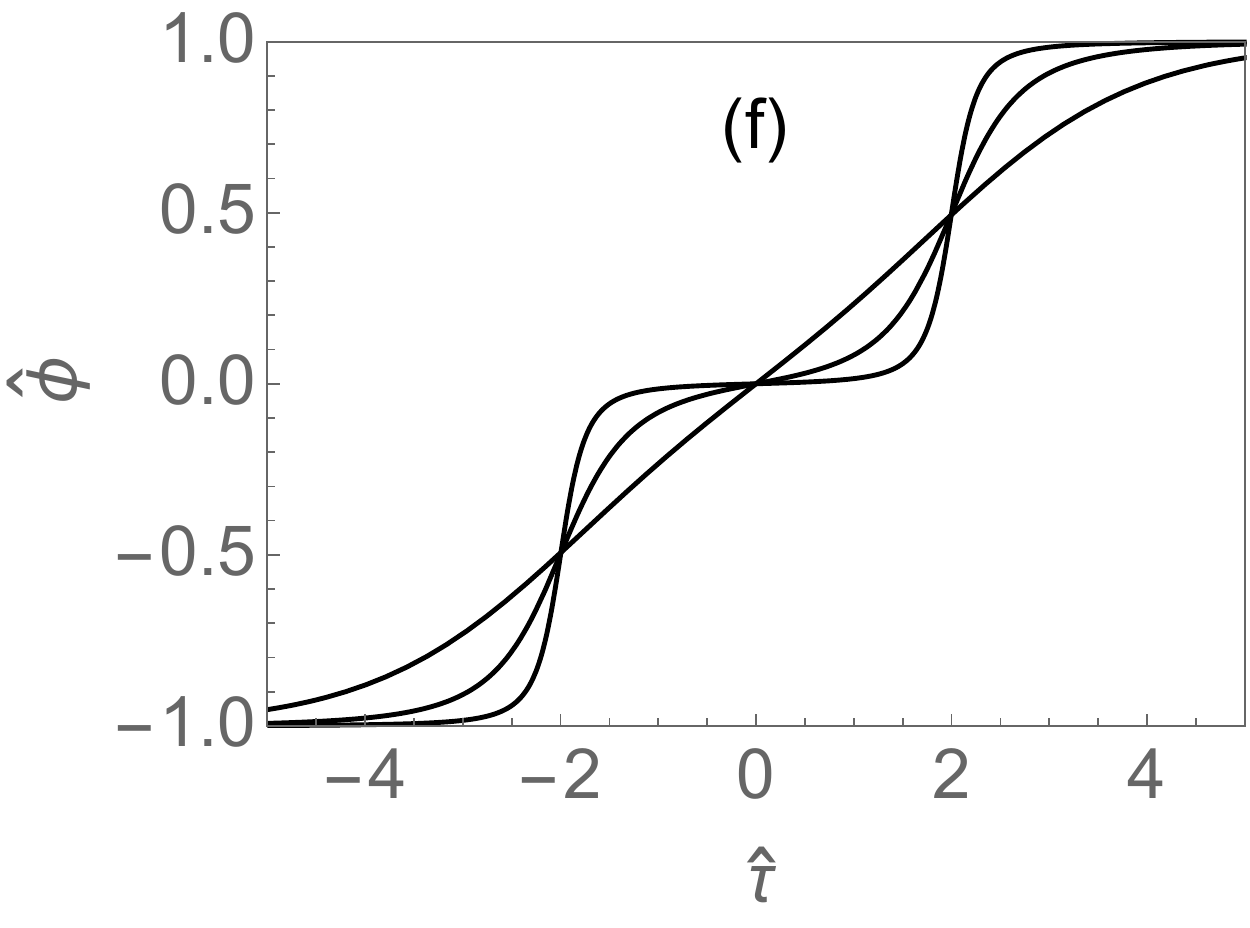}
\includegraphics[width=40mm]{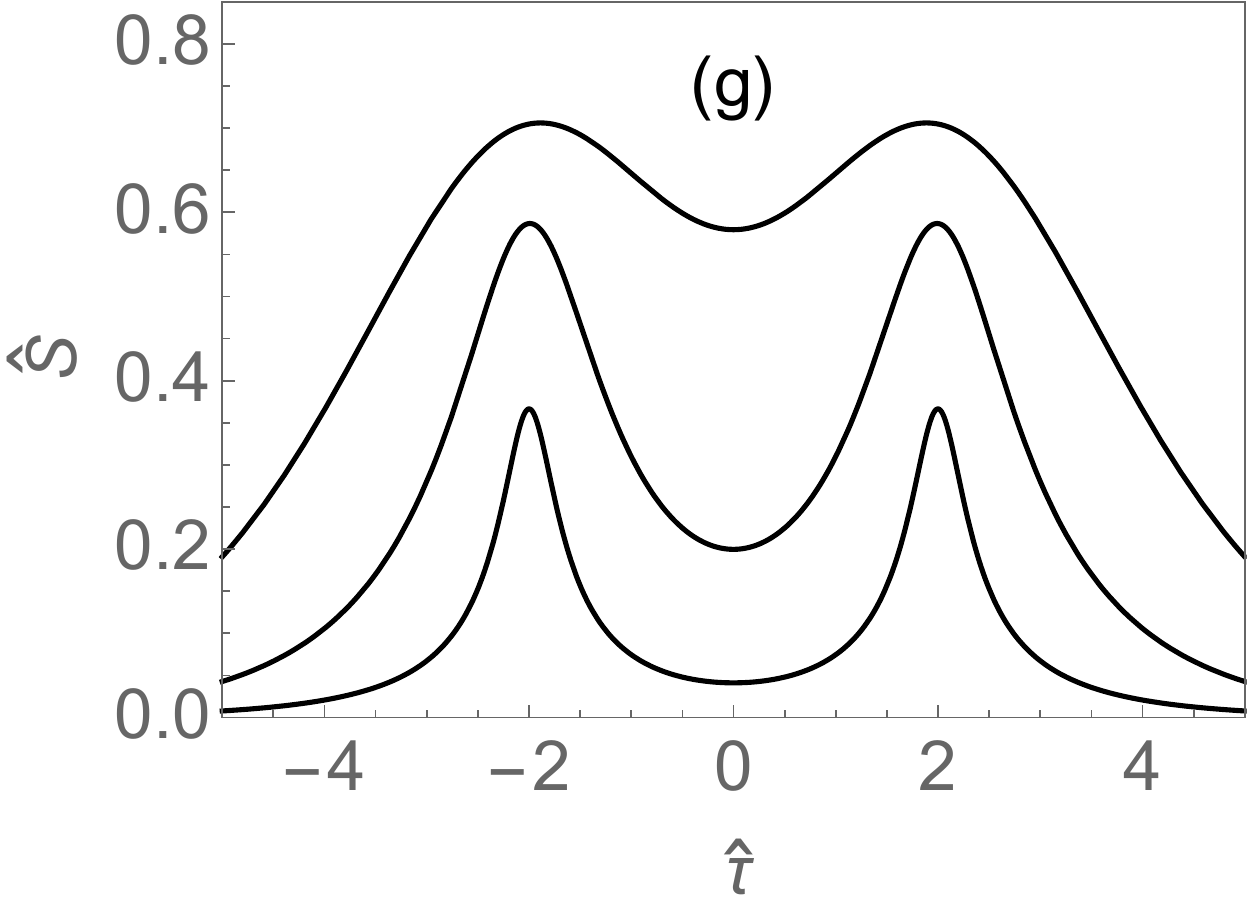}\hspace*{3mm}\includegraphics[width=40mm]{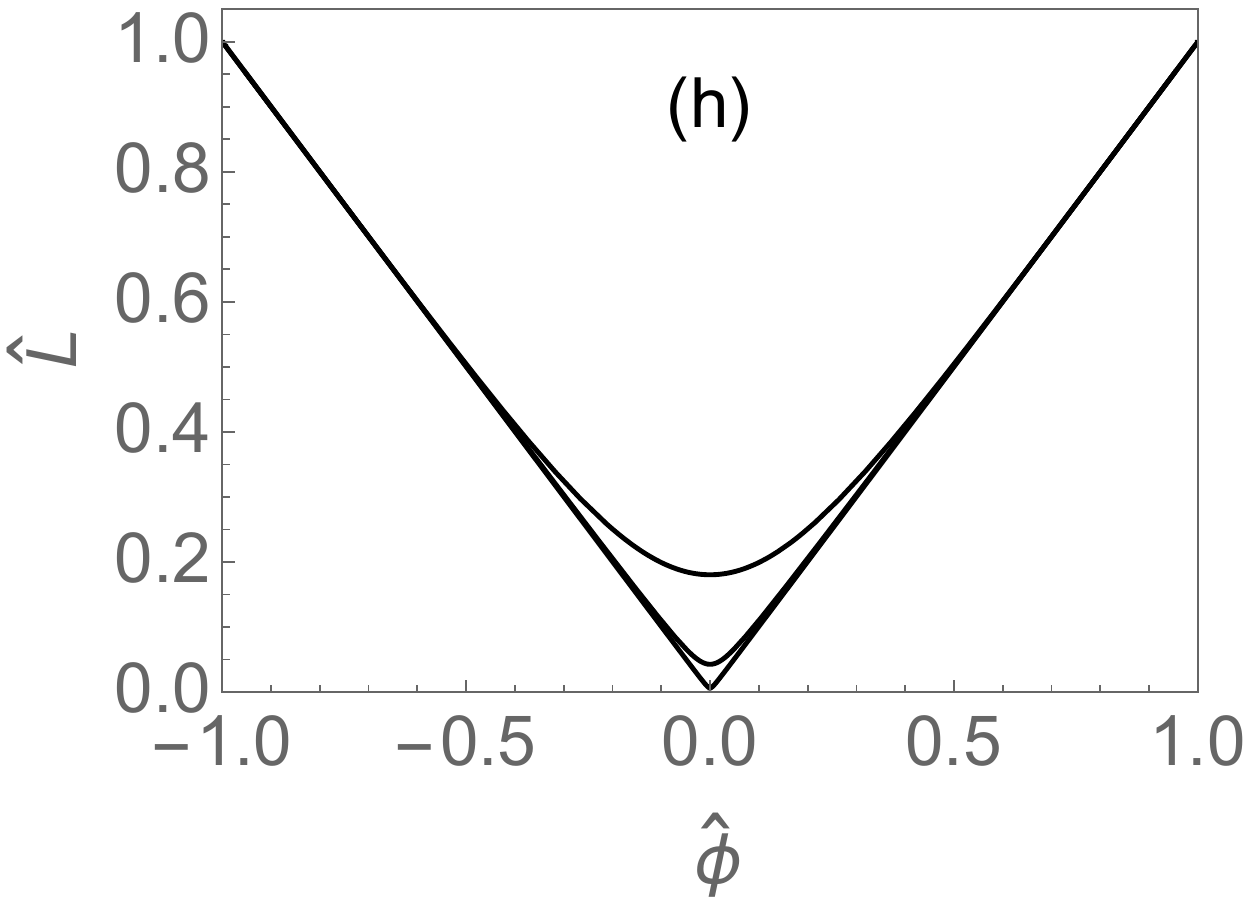}
\end{center}
\caption{Population densities of (a) host particles and (b) tag particles, both versus tension $\hat{J}$ and torque $\hat{\tau}$ at constant temperature $\hat{T}=1$ and zero cooperativity, $\Delta\gamma_\mathrm{t}=0$.
The nine contour lines from dark to bright in each panel are at (a) $0.025, \ldots, 0.225$, and (b) $0.096, \ldots, 0.864$. 
Panels (c)-(g) show the same two quantities plus contraction length $\hat{L}$, twist $\hat{\phi}$, and entropy $\hat{S}$ plotted versus $\hat{\tau}$ at constant tension $\hat{J}=1$ and temperature $\hat{T}=1$ for cooperativity with strength increasing from zero: $\Delta\gamma_\mathrm{t}/\gamma_\mathrm{t}=0,2,4$. Panel (h) shows contraction length versus twist angle.}
  \label{fig:figure15}
\end{figure}

In Fig.~\ref{fig:figure15} we present some explicit results.
The specifications in use are
\begin{align}\label{eq:28}
& L_1=L_2=L_3\doteq  L_\mathrm{t}, \nonumber \\
& \phi_1=\phi_2=-\phi_3=-\phi_4\doteq\phi_\mathrm{t}, \nonumber \\
& \gamma_2=\gamma_4\doteq\gamma_\mathrm{t}, \quad 
 \gamma_1=\gamma_3=\gamma_\mathrm{t}+\Delta\gamma_\mathrm{t}.
\end{align}
For the graphical presentation we adopt the following rescaling conventions:
\begin{align}\label{eq:29}
& \hat{L}\doteq\frac{|\bar{L}|}{L_\mathrm{t}},\quad
\hat{\phi}\doteq\frac{\bar{\phi}}{\phi_\mathrm{t}},\quad
\hat{S}\doteq\frac{\bar{S}}{k_\mathrm{B}}, \nonumber \\
& \hat{J}\doteq\frac{JL_\mathrm{t}}{\gamma_\mathrm{t}},\quad
\hat{\tau}\doteq\frac{\tau\phi_\mathrm{t}}{\gamma_\mathrm{t}},\quad
\hat{T}\doteq\frac{k_\mathrm{B}T}{\gamma_\mathrm{t}}.
\end{align}
 In panels (a) and (b) we show the population densities of hosts 1, 3 and tags 2, 4 versus torque and tension for the case of zero cooperativity and fixed temperature.
Panels (c)-(g) show the same population densities plus the thermodynamic functions $\hat{L}$, $\hat{\phi}$, $\hat{S}$, all versus torque at fixed tension, temperature (one value each), and cooperativity (three values). The dependence of contraction length on twist angle is shown in panel (h).

At fixed tension and with positive torque of increasing strength, twisted segments are being nucleated through the activation of hosts 1. 
Twisted segments grow in length through the activation of tags 2.
Adjacent segments of twisted chain merge as hosts are crowded out by tags.
Negative torque favors the activation and growth of hosts 3 and tags 4.
Any increase in the cooperativity is controlled by raising the nucleation threshold $\Delta\gamma_\mathrm{t}$ from zero.
Fewer extended segments are being nucleated.
However, once they are nucleated, they grow more rapidly with torque of increasing strength.
The changes in contraction length and twist angle take place more abruptly.
The entropy is lower overall and peaks when the torque crosses the nucleation threshold.

Note the similarities and differences between temperature effects in Fig.~\ref{fig:figure5}(d)-(f) and Fig.~\ref{fig:figure15}(e)-(g).
For the quantitative analysis of transformations between conformations in DNA under tension and torque, effects of cooperativity are important.
It takes nested particles to model cooperativity in an natural way.
The dependence of contraction length $\hat{L}$ on $\hat{\phi}$ as depicted in Fig.~\ref{fig:figure15}(h) is a key quantity, directly accessible to experiments on DNA. 
Here we see what twist alone produces. 
This simple, near linear dependence will be modified by the presence of supercoiling in some settings.

\subsection{High torque: twisted and supercoiled chain}\label{sec:hig-tor-reg}
\begin{figure}[htb]
  \begin{center}
\includegraphics[width=85mm]{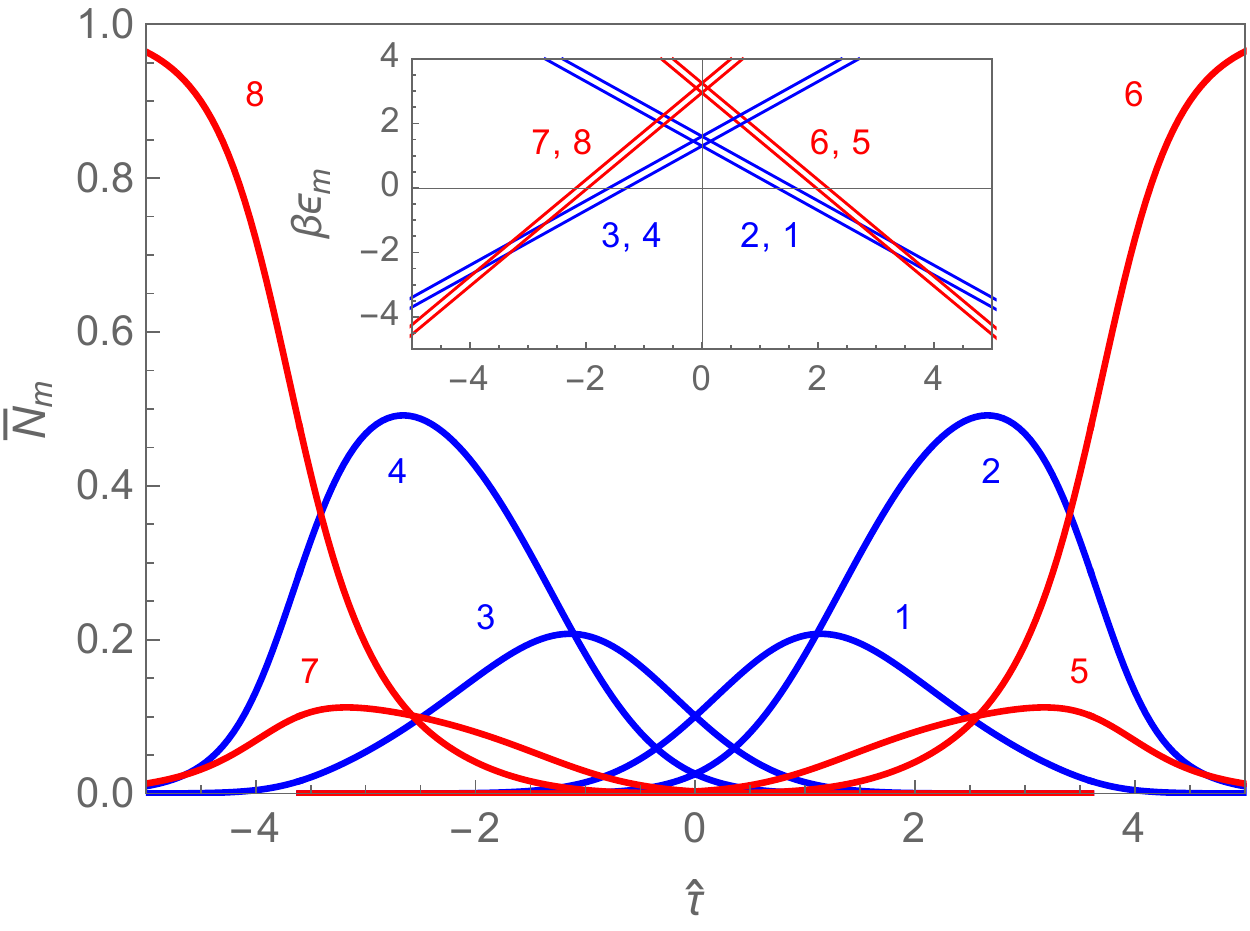}
\end{center}
\caption{Population densities of twist particles 1, 3 (hosts), 2, 4 (tags), and supercoil particles 5, 7 (hybrids), 6, 8 (tags) versus torque $\hat{\tau}$ at constant tension $\hat{J}$ and temperature $\hat{T}$ with specifications (\ref{eq:30a}), (\ref{eq:30b}). 
The inset shows the activation energies $\beta\epsilon_m$ versus torque $\hat{\tau}$.}
  \label{fig:figure16}
\end{figure}

At sufficiently strong (positive) torque, the activation energies of supercoil particles 5,6 descend below those of twist particles 1,2 as illustrated in the inset to Fig.~\ref{fig:figure16} for one representative case with the following specifications:
\begin{align}\label{eq:30a}
& \gamma_1=\gamma_3\doteq\gamma_\mathrm{t}+\Delta\gamma_\mathrm{t}, \quad
\gamma_2=\gamma_4\doteq\gamma_\mathrm{t}, \nonumber \\
& \gamma_5=\gamma_7\doteq\gamma_\mathrm{s}+\Delta\gamma_\mathrm{s}, \quad
\gamma_6=\gamma_8\doteq\gamma_\mathrm{s}, \nonumber \\
& L_1=L_2=L_3=L_4\doteq L_\mathrm{t}, \quad
L_5=L_6=L_7=L_8\doteq L_\mathrm{s}, \nonumber \\
& \phi_1=\phi_2=-\phi_3=-\phi_4\doteq \phi_\mathrm{t}, \nonumber \\
& \phi_5=\phi_6=-\phi_7=-\phi_8\doteq \phi_\mathrm{s}, 
\end{align}
\begin{align}\label{eq:30b}
& \beta\gamma_\mathrm{t}=1, \quad
\gamma_\mathrm{s}/\gamma_\mathrm{t}=2.5, \quad
 \Delta\gamma_\mathrm{t}/\gamma_\mathrm{t}=0.3, \quad
\Delta\gamma_\mathrm{s}/\gamma_\mathrm{t}=0.3, \nonumber \\
& JL_\mathrm{t}/\gamma_\mathrm{t}=0.3, \quad
JL_\mathrm{s}/\gamma_\mathrm{t}=0.45, \quad
\phi_\mathrm{s}/\phi_\mathrm{t}=1.5.
\end{align}
At constant tension, the $\epsilon_m$ vary linearly with torque.
Positive (negative) torque lowers $\epsilon_m$ for $m=1,2,5,6$ ($m=3,4,7,8$).
By design, the slope is steeper and the intercept higher for supercoil particles, $m=5,6,7,8$, than for twist particles, $m=1,2,3,4$.
Cooperativity for twist (supercoil) particles displaces the line for hosts 1,3 (hybrids 5,7) upward relative to the line for tags 2,4 (6,8).

The statistical mechanical analysis of the general case with all eight particles in play now requires that we solve the nonlinear Eqs.~(\ref{eq:5}) and the linear Eqs.~(\ref{eq:6}) for all eight species.
The former reduce to a fifth-order polynomial equation with a unique physical solution.
In the main plot of Fig.~\ref{fig:figure16} we show the population density $\bar{N}_m$ for each particle species versus torque at fixed tension and temperature.
It pertains to a tension of moderate strength and a response of moderate cooperativity for both twist and supercoil particles.
The temperature selected is also of intermediate value.

At zero torque only hosts 1 and 3 have significant populations.
They represent very short segments of twisted chain, activated by thermal fluctuations with no bias in orientation.
A weak positive torque enhances such fluctuations of one orientation (hosts 1) at the expense of the other (hosts 3).
With increasing (positive) torque the twist segments nucleated by hosts 1 grow in size through the activation of an increasing number of tags 2.
As the segments of twisted chain grow in size they begin to merge, which reduces their number. 
This process is reflected in a decrease of the density of hosts, $\bar{N}_1$, as the density of tags, $\bar{N}_2$, continues to increase.

The threshold for the appearance of supercoil particles is reached when their activation energies descend below those of the twist particles (see inset).
The sharpness of this threshold depends on cooperativity.
At this threshold, supercoil segments begin to nucleate.
The mechanism of this nucleation is the activation of hybrids 5 and the mechanism for the growth of supercoil segments is the activation of tags 6.
Under yet stronger torque the supercoil particles begin to crowd out twist particles, which means that the fraction of supercoiled chain increases at the expense of merely twisted chain.
As the supercoil segments grow, they begin to merge just as the twisted segments did at weaker torque.
This is reflected in the decrease of $\bar{N}_5$ (hybrids) and increase of $\bar{N}_6$ (tags).
At very strong torque, the system is in the conformation of a single supercoil segment, composed of one host 1, one hybrid 5, and a macroscopic number of tags 6.

\begin{figure}[htb]
  \begin{center}
\includegraphics[width=40mm]{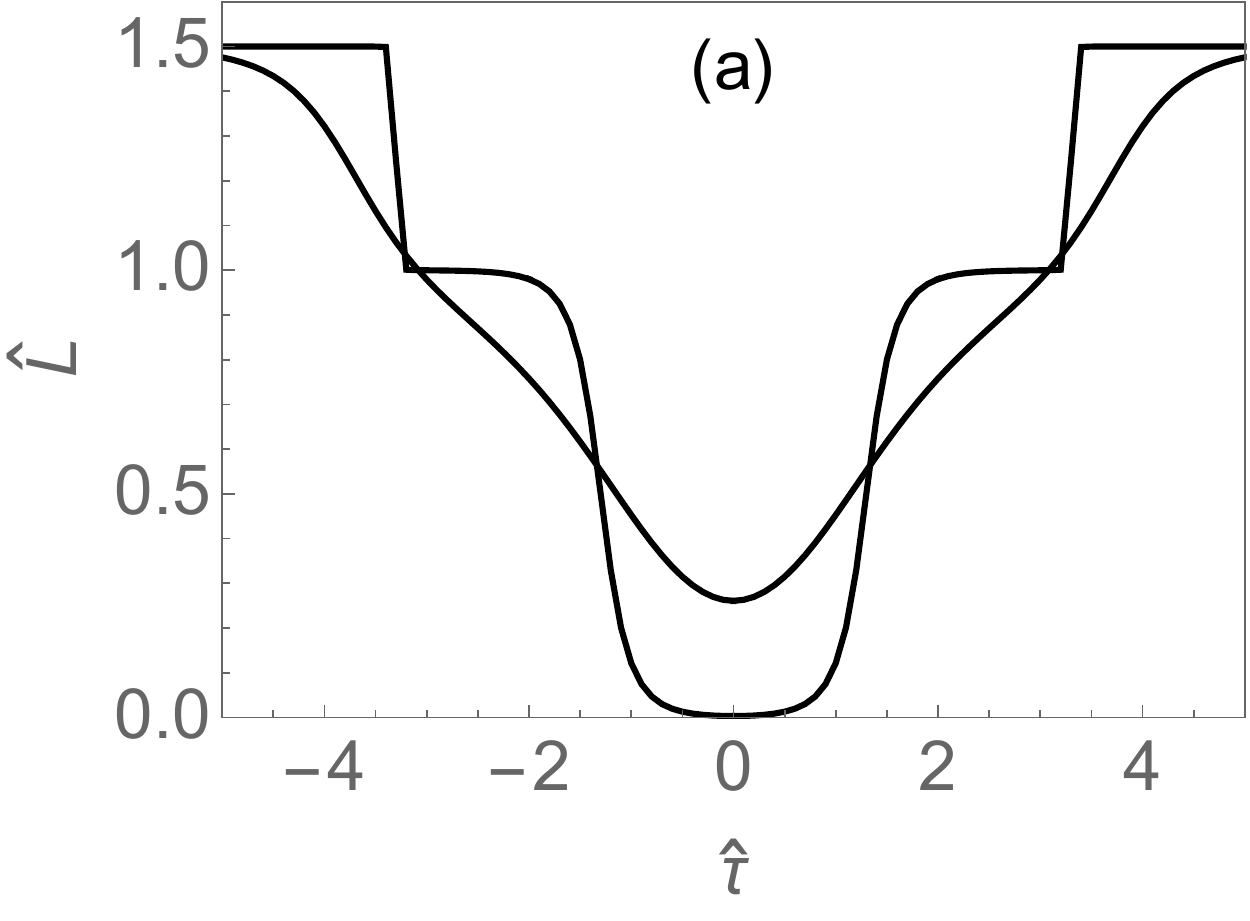}\hspace*{3mm}\includegraphics[width=40mm]{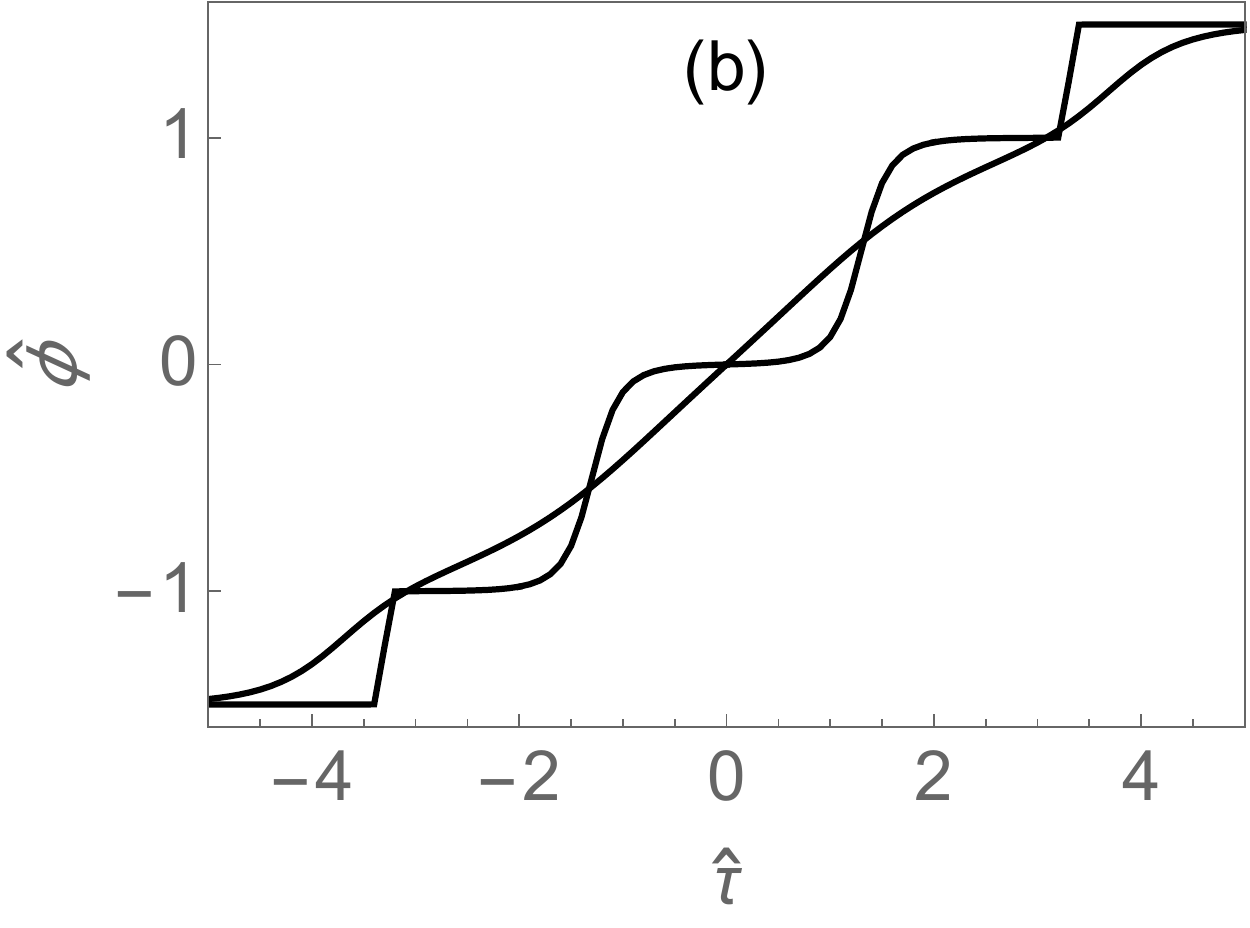}
\includegraphics[width=40mm]{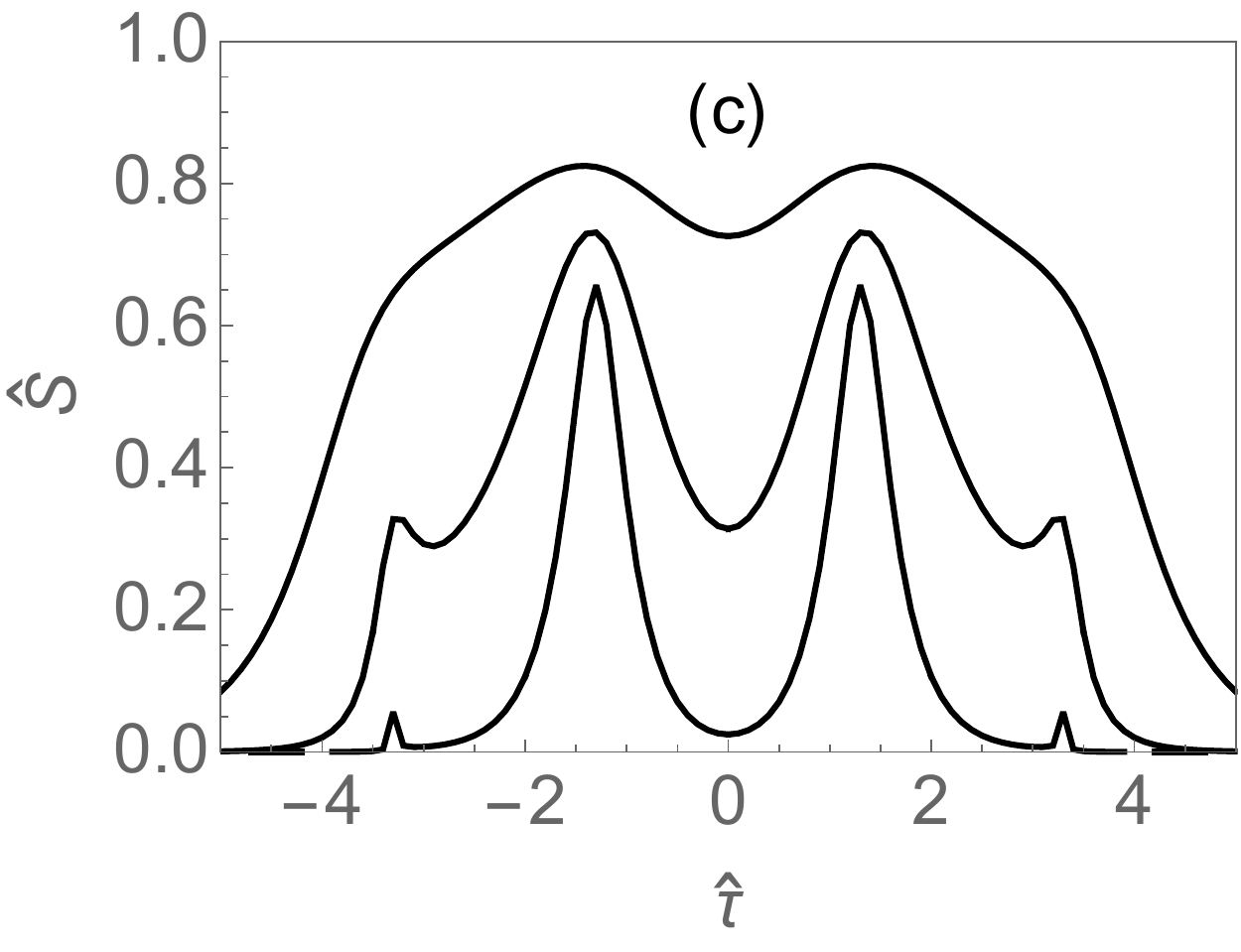}
\includegraphics[width=40mm]{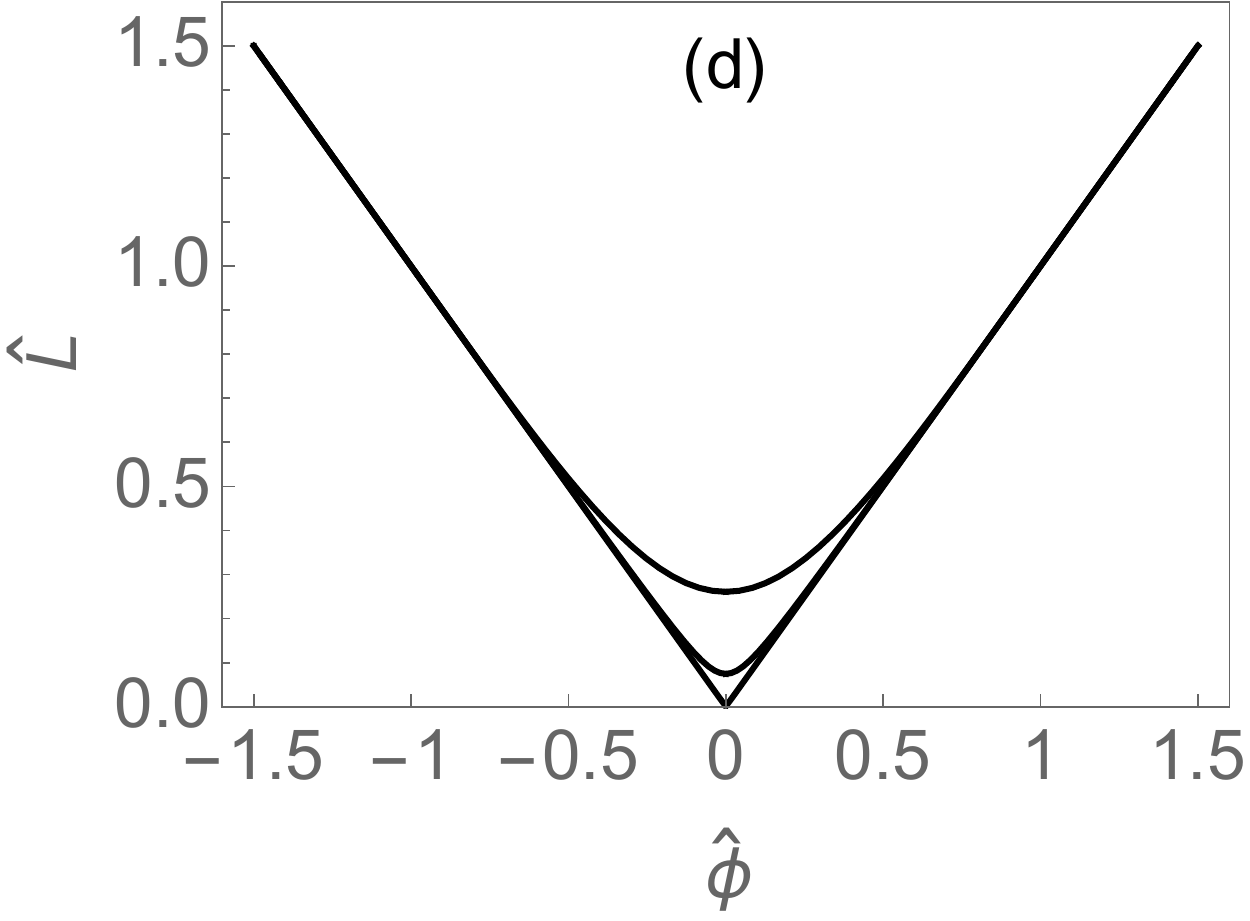}
\end{center}
\caption{(a) Contraction distance $\hat{L}$, (b) linkage $\hat{\phi}$, and (c) entropy $\hat{S}$, all versus torque $\hat{\tau}$ at constant tension $\hat{J}$ and temperature $\hat{T}$.
Panel (d) shows $\hat{L}$ versus $\hat{\phi}$.
The three curves in (c) are for $\hat{T}=1, 0.5, 0.25$. 
Data for intermediate $\hat{T}$ are not shown in (a) and (b). 
The remaining specifications are from (\ref{eq:30a}), (\ref{eq:30b}).}
  \label{fig:figure18}
\end{figure}

The effects of supercoil formation on length contraction, linkage angle, and entropy are shown in Fig.~\ref{fig:figure18}.
The characteristic features associated with supercoils stand out in a comparison of data at two or three different temperatures.
These features are especially prominent when Figs.~\ref{fig:figure18}(a)-(c) are compared with Figs.~\ref{fig:figure15}(e)-(g), where the supercoil response is suppressed.
From Fig.~\ref{fig:figure16} we know that the activation of supercoils sets in at $\hat{\tau}\gtrsim 3.5$.
The signature of supercoil formation is an enhancement in linkage at the cost of an augmented contraction.
Both effects are barely visible at the highest $\hat{T}$ but very clearly at the lowest $\hat{T}$.

The entropy curve at the highest $\hat{T}$ has a broad maximum for intermediate torque.
This reflects strong fluctuations involving competing conformations.
All of them are represented by particles with activation energies near zero.
Only at high torque does the supercoil conformation become predominant and particles 6 crowd out all others, which is reflected in a pronounced entropy drop.
At the intermediate temperature, the signature of supercoil formation becomes discernible in the entropy data as a shoulder-like feature at $\hat{\tau}\simeq3.5$.
The sharpened maximum (at $\hat{\tau}\simeq1.5$) marks the torque threshold for the activation of twisted segments.
The features associated with twisted chain are similar to what we have already described in the context of Fig.~\ref{fig:figure15}.
The appearance of supercoils manifests itself by way of sharp increases in contraction length and linkage as well as by a sharp spike in the entropy.

\subsection{Showcase for particle nesting}\label{sec:sho-cas-nes}
In conclusion of Sec.~\ref{sec:sup-coi-ht} with focus on high-tension supercoils and in anticipation of Sec.~\ref{sec:sup-coi-lt}, where our focus will shift to low-tension supercoils, we are well positioned to highlight the role of particle nesting within the framework of statistically interacting particles.
Particle nesting is the key feature in the statistical mechanical analysis of this rather complex behavioral trait manifest in stretched and twisted polymeric chains.

The same set of eight particle species, introduced in Sec.~\ref{sec:sup-coi} and consisting of two hierarchies, one for each sense of chirality, can be used to describe, in flexible polymers, the appearance of supercoils out of highly twisted chains and, with different parameter settings for stiff polymers, the formation of supercoils out of largely untwisted chains.
The nesting structure, which is determined by the combinatorial specifications given in Table~\ref{tab:2}, is identical for low-tension and high-tension supercoils even though some particle species take on very different roles in the two applications.  
The specific role is determined by energetic parameters.

The same two hierarchies thus describe, in one parameter setting, how a twisted chain converts into a supercoiled chain under increasing torque at fixed tension (shown earlier in Sec.~\ref{sec:sup-coi-ht}) and, in a different parameter setting, how supercoiled chain gives way to twisted chain when torque is increased at fixed tension (to be demonstrated in Sec.~\ref{sec:sup-coi-lt}).
With some modifications this model has applications for ds-DNA, which we plan to explore in continuation of this project \cite{mct3}.

%
\section{Supercoils at low tension}\label{sec:sup-coi-lt}
\begin{figure}[htb]
  \begin{center}
\includegraphics[width=85mm]{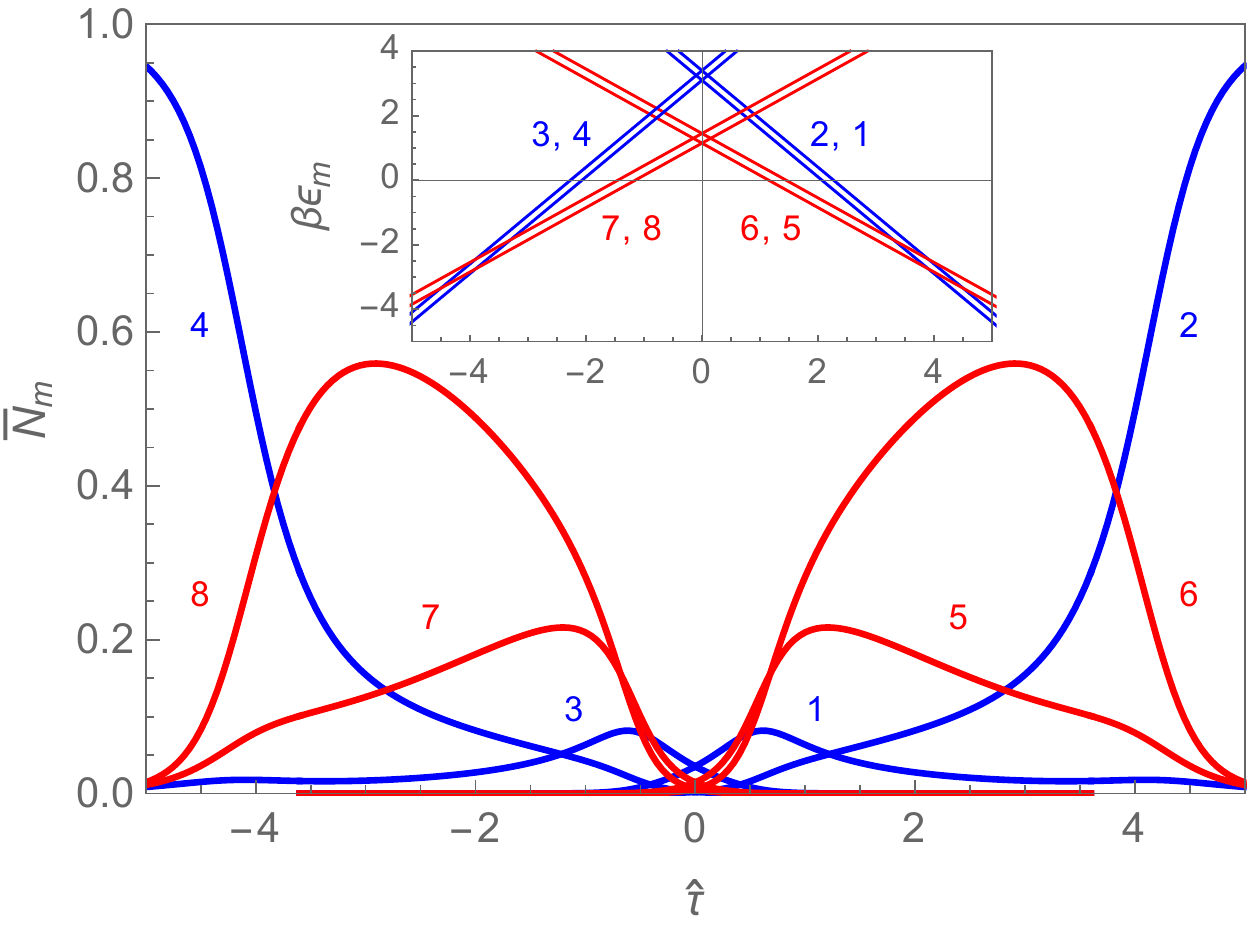}
\end{center}
\caption{Population densities of twist particles 1, 3 (hosts), 2, 4 (tags), and supercoil particles 5, 7 (hybrids), 6, 8 (tags)  versus  torque $\hat{\tau}$ at constant tension $\hat{J}$ and temperature $\hat{T}$. The inset shows the activation energies $\beta\epsilon_m$ versus torque $\hat{\tau}$ for the case with specifications (\ref{eq:32}).}
  \label{fig:figure24}
\end{figure}

%
In molecular chains of strong torsional stiffness, supercoils tend to form out of segments that are weakly twisted at most, within a regime of low tension and not too high torque.
It then suffices that we reverse the inequalities (\ref{eq:20b}) and (\ref{eq:20c}) into 
\begin{align}\label{eq:24} 
& \phi_5<\phi_1,\quad |\phi_7|<|\phi_3|,\quad \phi_6<\phi_2,\quad |\phi_8|<|\phi_4|, \nonumber \\ 
& \gamma_5<\gamma_1,\quad \gamma_7<\gamma_2,\quad \gamma_6<\gamma_2,\quad \gamma_8<\gamma_4,
\end{align}
while leaving inequalities (\ref{eq:20a}) intact.
In consequence, the dependence on torque of the activation energies for twist particles and supercoil particles is interchanged (compare insets to Figs.~\ref{fig:figure24} and \ref{fig:figure16}).
We change four of the six scaling conventions (\ref{eq:29}) for this section as follows:
\begin{align}\label{eq:31}
\hat{\phi}\doteq\frac{\bar{\phi}}{\phi_\mathrm{s}},\quad
 \hat{J}\doteq\frac{JL_\mathrm{t}}{\gamma_\mathrm{s}},\quad
\hat{\tau}\doteq\frac{\tau\phi_\mathrm{t}}{\gamma_\mathrm{s}},\quad
\hat{T}\doteq\frac{k_\mathrm{B}T}{\gamma_\mathrm{s}}.
\end{align}
The energetic specifications for the case analyzed in the following are (\ref{eq:30a}) and
\begin{align}\label{eq:32}
& \beta\gamma_\mathrm{s}=1, \quad
\gamma_\mathrm{t}/\gamma_\mathrm{s}=3.0, \quad
 \Delta\gamma_\mathrm{t}/\gamma_\mathrm{s}=0.3, \quad
\Delta\gamma_\mathrm{s}/\gamma_\mathrm{s}=0.3, \nonumber \\
& JL_\mathrm{t}/\gamma_\mathrm{s}=0.1, \quad
JL_\mathrm{s}/\gamma_\mathrm{s}=0.15, \quad
\phi_\mathrm{t}/\phi_\mathrm{s}=1.5.
\end{align}

At zero torque, hosts 1 and 3, which nucleate segments of twisted chain, still have the highest populations of all particles, now greatly reduced. 
The next highest population densities pertain to hybrids 5 and 7.
Unlike in Sec.~\ref{sec:sup-coi-ht}, weak positive (negative) torque does not grow twisted segments by activating tags 2 (tags 4).
Instead it nucleates supercoil segments via the activation of hybrids 5 (hybrids 7) and grows with tags 6 (tags 8). 

As the torque continues to gain strength, the supercoiled segments begin to merge.
The signature of that effect is a decrease in the populations of hybrids 5, (hybrids 7) combined with an increase in the populations of tags 6 (tags 8).
Only at yet stronger torque is the torsional stiffness of the chain overcome.
Supercoiled segments now convert into segments of twisted chain of increasing size. 
The population of tags 6 (tags 8) collapses while the population of tags 2 (tags 4) shoots up.

\begin{figure}[b]
  \begin{center}
\includegraphics[width=40mm]{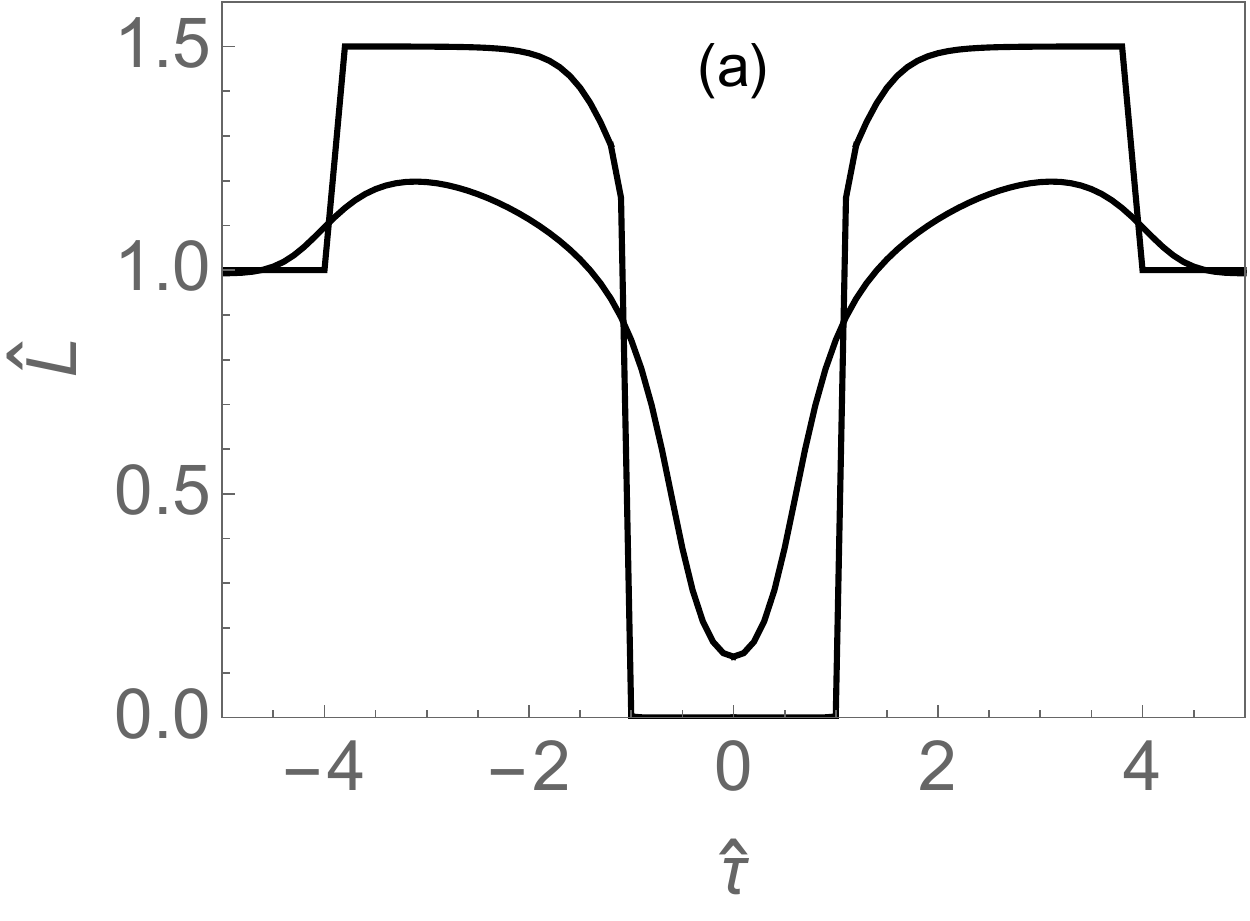}\hspace*{3mm}\includegraphics[width=40mm]{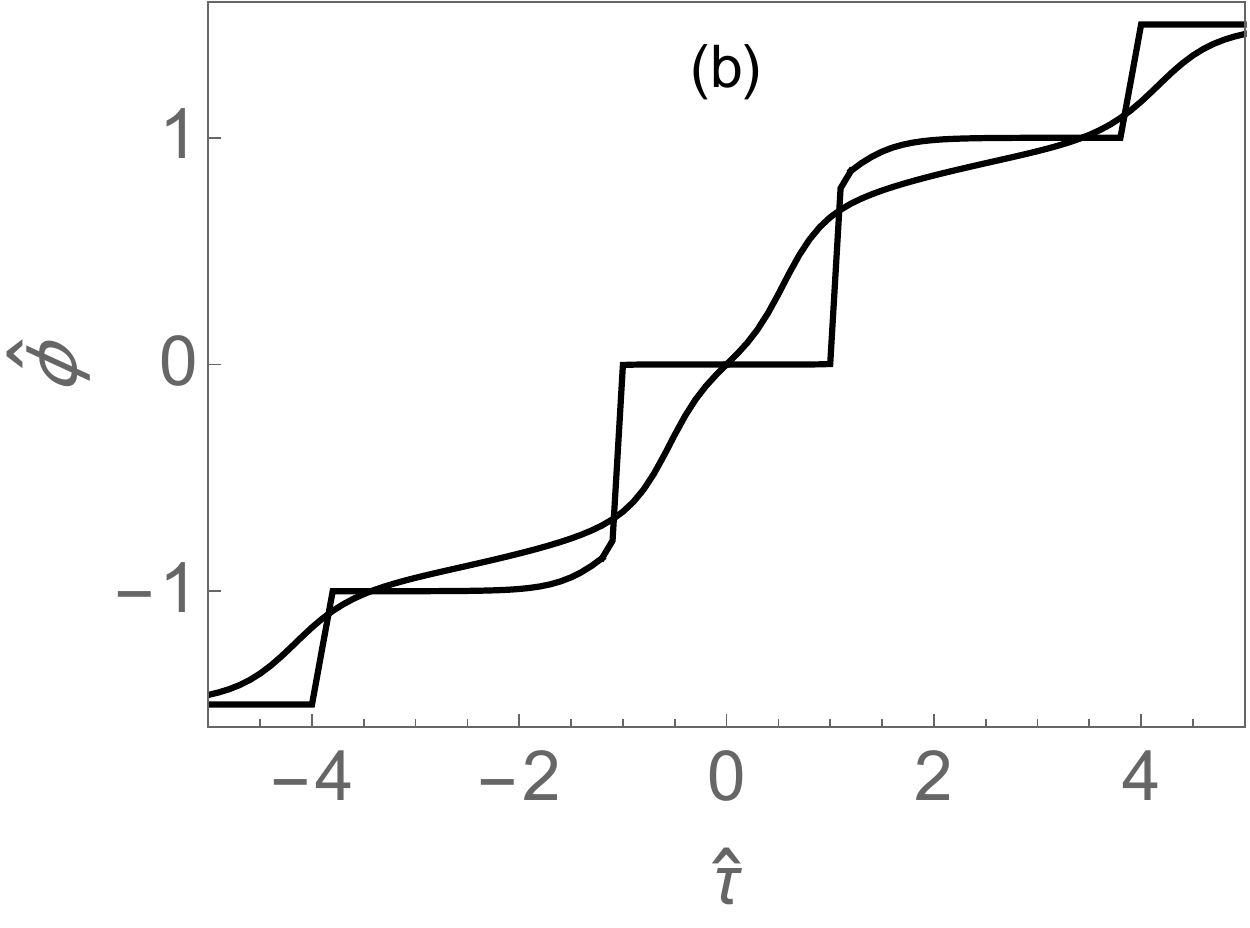}
\includegraphics[width=40mm]{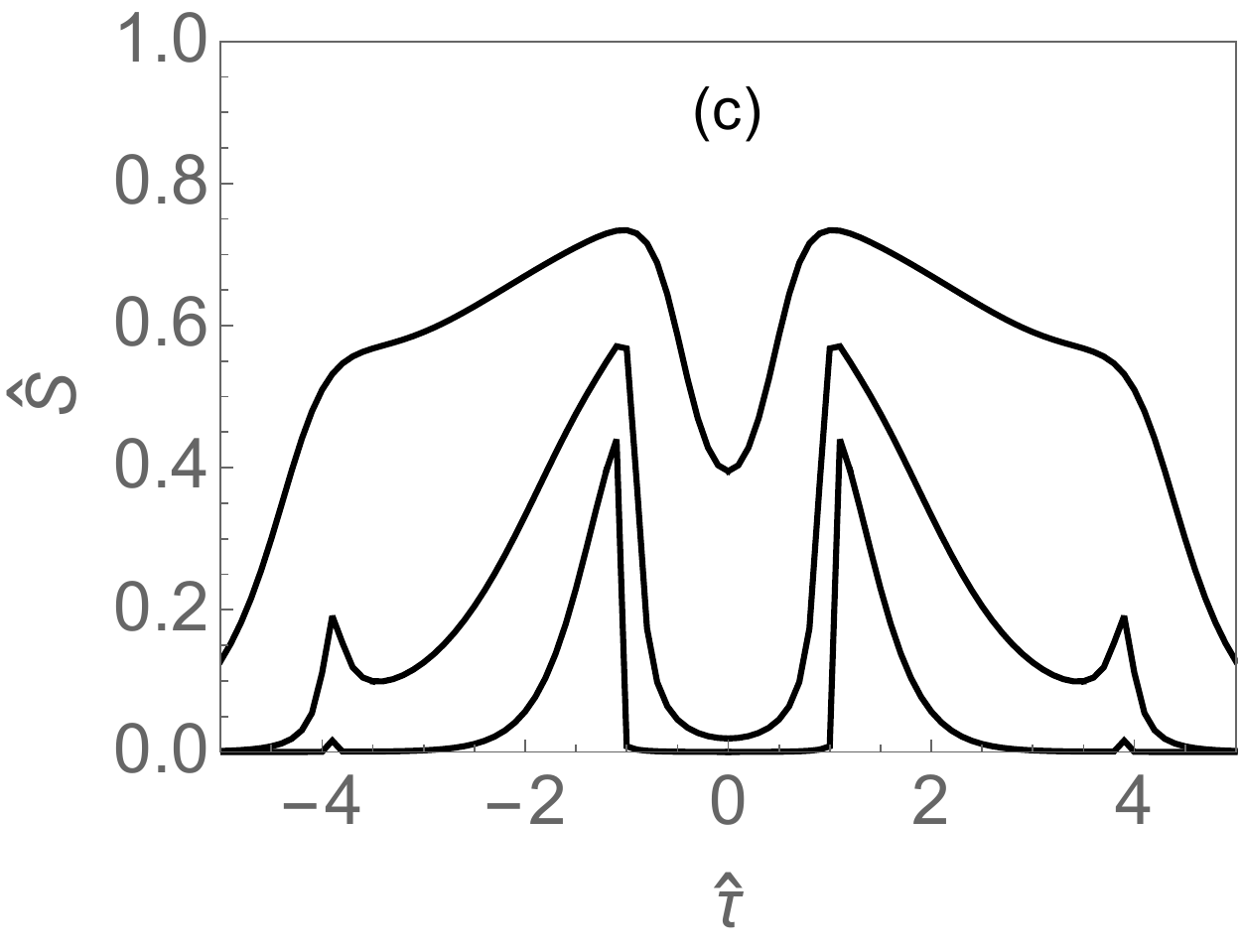}\hspace*{3mm}\includegraphics[width=40mm]{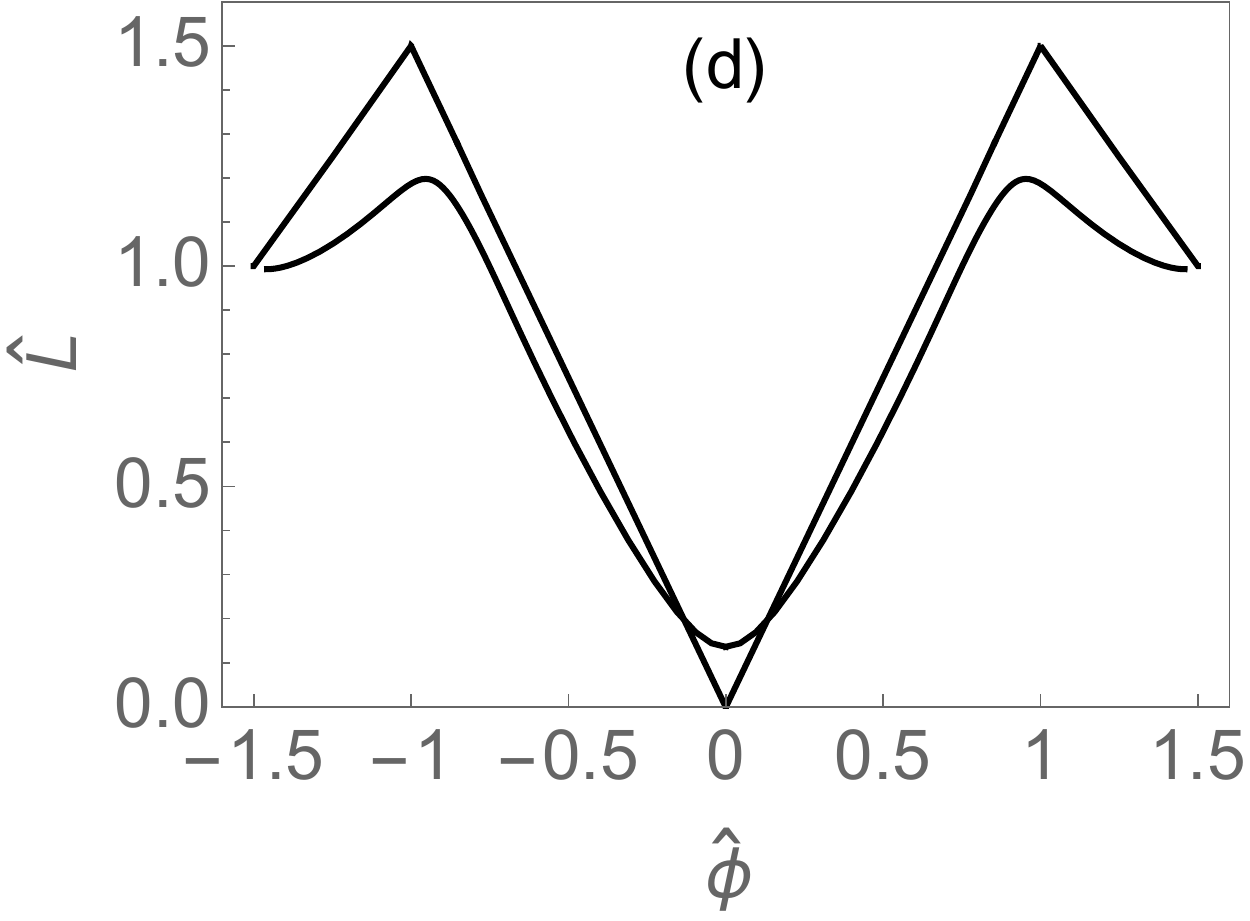}
\end{center}
\caption{(a) Contraction distance $\hat{L}$, (b) linkage $\hat{\phi}$, and (c) entropy $\hat{S}$, all versus torque $\hat{\tau}$ at constant tension $\hat{J}$ and temperature $\hat{T}$.
Panel (d) shows $\hat{L}$ versus $\hat{\phi}$.
The three curves in (c) are for $\hat{T}=1, 0.5, 0.25$. 
Data for the intermediate temperature are not shown in (a), (b), and (d). 
The remaining specifications are from (\ref{eq:30a}), (\ref{eq:32}).}
  \label{fig:figure25}
\end{figure}

The effects on length contraction, linkage, and entropy of the formation of low-tension supercoils at weak torque and their conversion into twisted chain at stronger torque are shown in Fig.~\ref{fig:figure25}.
Some of the contrasts to the corresponding plots in Fig.~\ref{fig:figure18} are striking.
In each panel we can discern the following three major processes: the nucleation of supercoil segments at $|\hat{\tau}|\simeq1$, the merging of supercoil segments at $2\lesssim|\hat{\tau}|\lesssim3.5$, and their conversion into long twisted segments at $|\hat{\tau}|\simeq4$.

The nucleation of supercoil segments is gradual at the highest temperature and becomes rather precipitous at the lowest temperature, graphically represented, respectively, by soft and sharp increases in length contraction, linkage, and entropy. 
The merging of supercoil segments is associated with small increases in length contraction and linkage, but a marked decrease in entropy, especially at the lower temperatures.
The conversion of supercoiled chain into twisted chain combines an increase in linkage with a decrease in contraction length, which is gradual at the highest $T$ and abrupt at the lowest $T$. 
The effect on entropy is a spike, which is small but prominent at low $T$. 
With increasing $T$, this signal gets more and more washed out by thermal fluctuations into a shoulder-like feature.

%
\section{Buckling transition}\label{sec:DNA-plect}
%
One intensively investigated phenomenon in single-molecule experiments on ds-DNA under tension and torque is the buckling transition, specifically the rather abrupt formation of plectonemes when the twist angle is gradually advanced under constant low tension \cite{SAB+96, Mark98, ABLC98, SABC98, HYZ99, SAB+99, BM00, ZZY00, SABC00, SACB00, BBS03, Neuk04, Mark07, FDS+08, DFS+09, SW09, MN13, ELS13}.
Here we demonstrate that the system of eight species of statistically interacting particles portrayed in Secs.~\ref{sec:sup-coi-ht} does indeed include a parameter regime that features a buckling transition. 

We return to the scaling convention (\ref{eq:29}) with a single modification,
\begin{align}\label{eq:33}
\hat{L}\doteq\frac{|\bar{L}|}{L_\mathrm{s}}.
\end{align}
We leave the specifications (\ref{eq:30a}) unchanged and modify 
the specifications (\ref{eq:30b}) to
\begin{align}\label{eq:34}
& \beta\gamma_\mathrm{t}=1, \quad
\gamma_\mathrm{s}/\gamma_\mathrm{t}=2.5, \quad
 \Delta\gamma_\mathrm{t}/\gamma_\mathrm{t}=0.3, \quad
\Delta\gamma_\mathrm{s}/\gamma_\mathrm{t}=0.3, \nonumber \\
& JL_\mathrm{t}/\gamma_\mathrm{t}=0.03, \quad
JL_\mathrm{s}/\gamma_\mathrm{t}=0.3, \quad
\phi_\mathrm{s}/\phi_\mathrm{t}=1.5.
\end{align}
These modifications have little impact on the population densities found for the case analyzed in Sec.~\ref{sec:hig-tor-reg} and shown in Fig.~\ref{fig:figure16}. 
The largest difference (not shown) is an enhancement in the population of twist tags 2 and 4.

\begin{figure}[t]
  \begin{center}
\includegraphics[width=40mm]{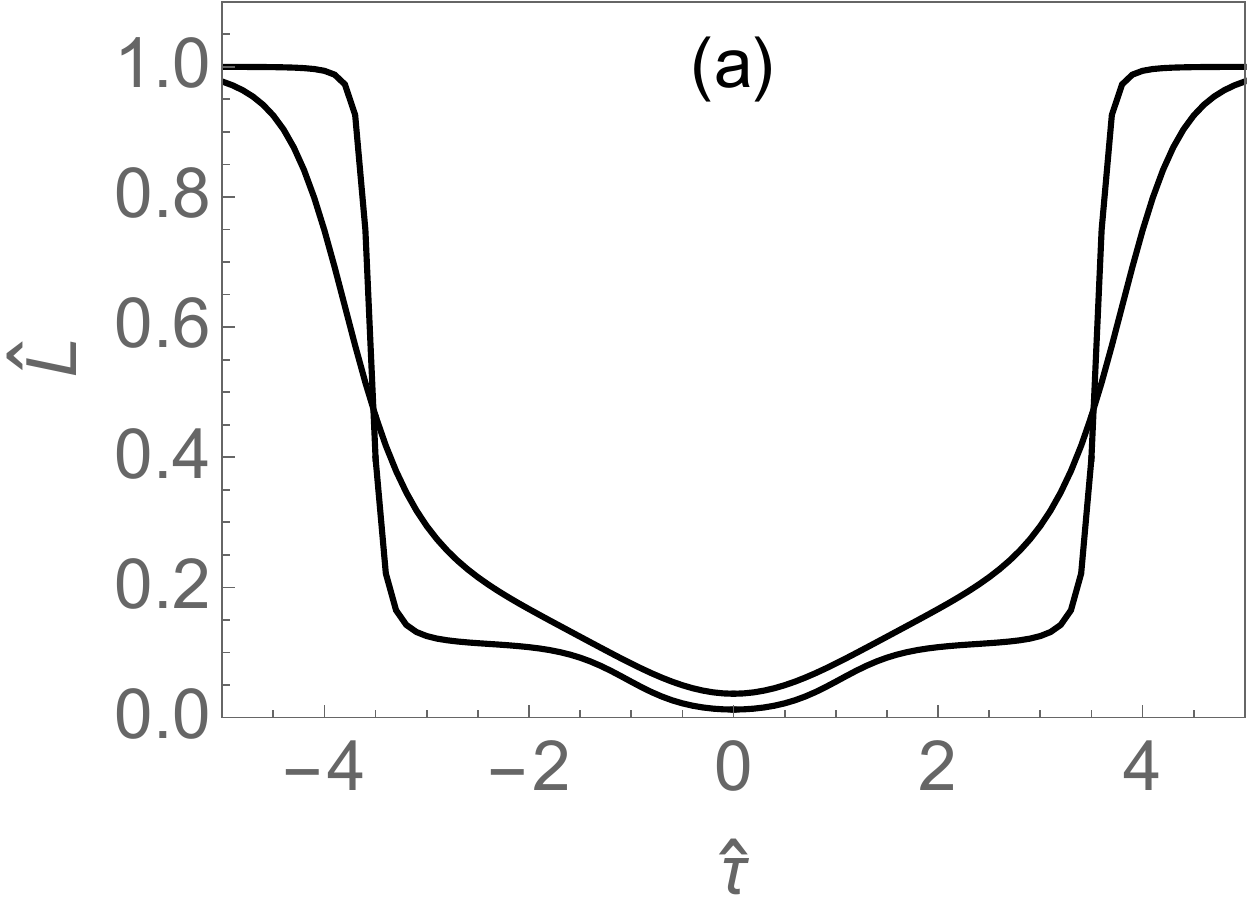}\hspace*{3mm}\includegraphics[width=40mm]{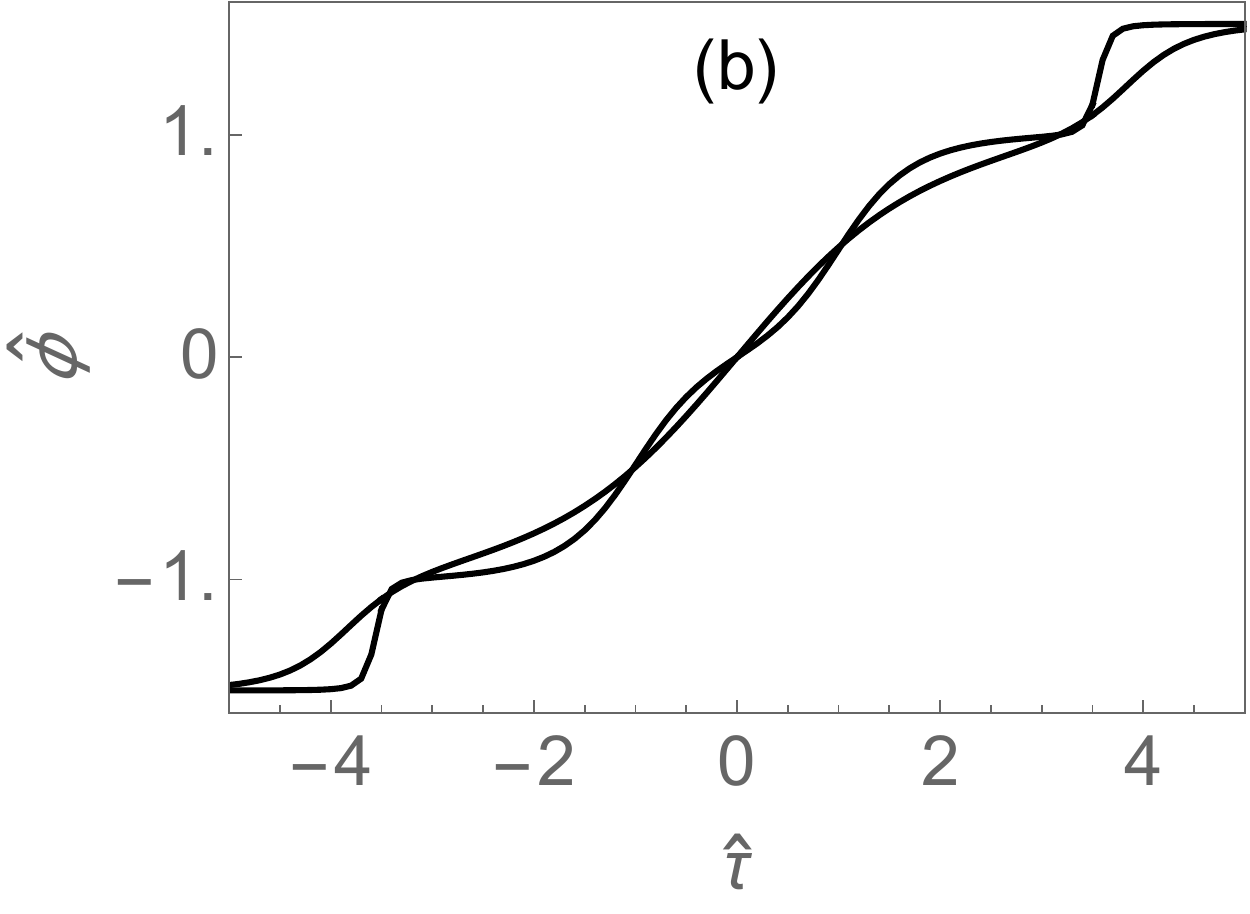}
\includegraphics[width=40mm]{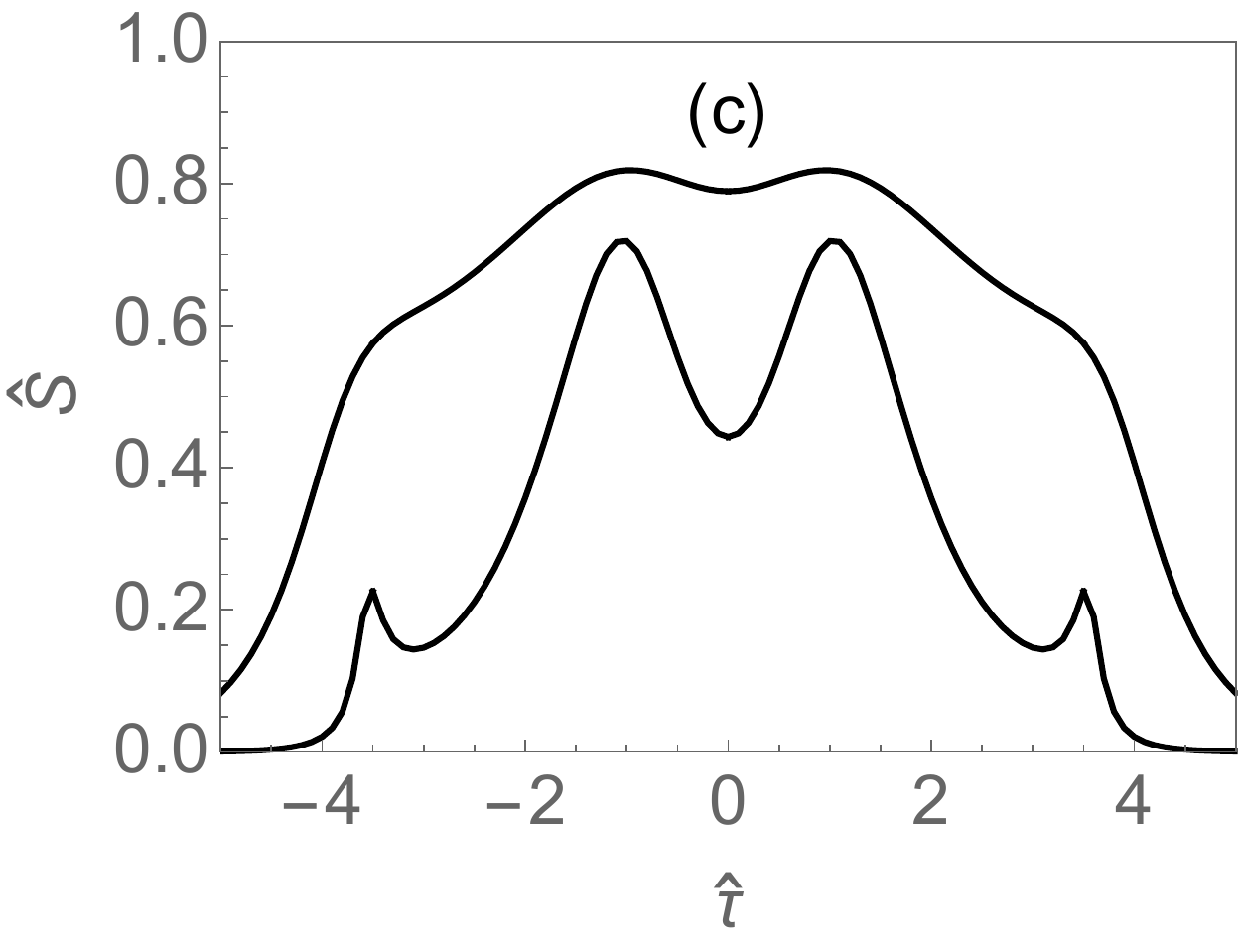}\hspace*{3mm}\includegraphics[width=40mm]{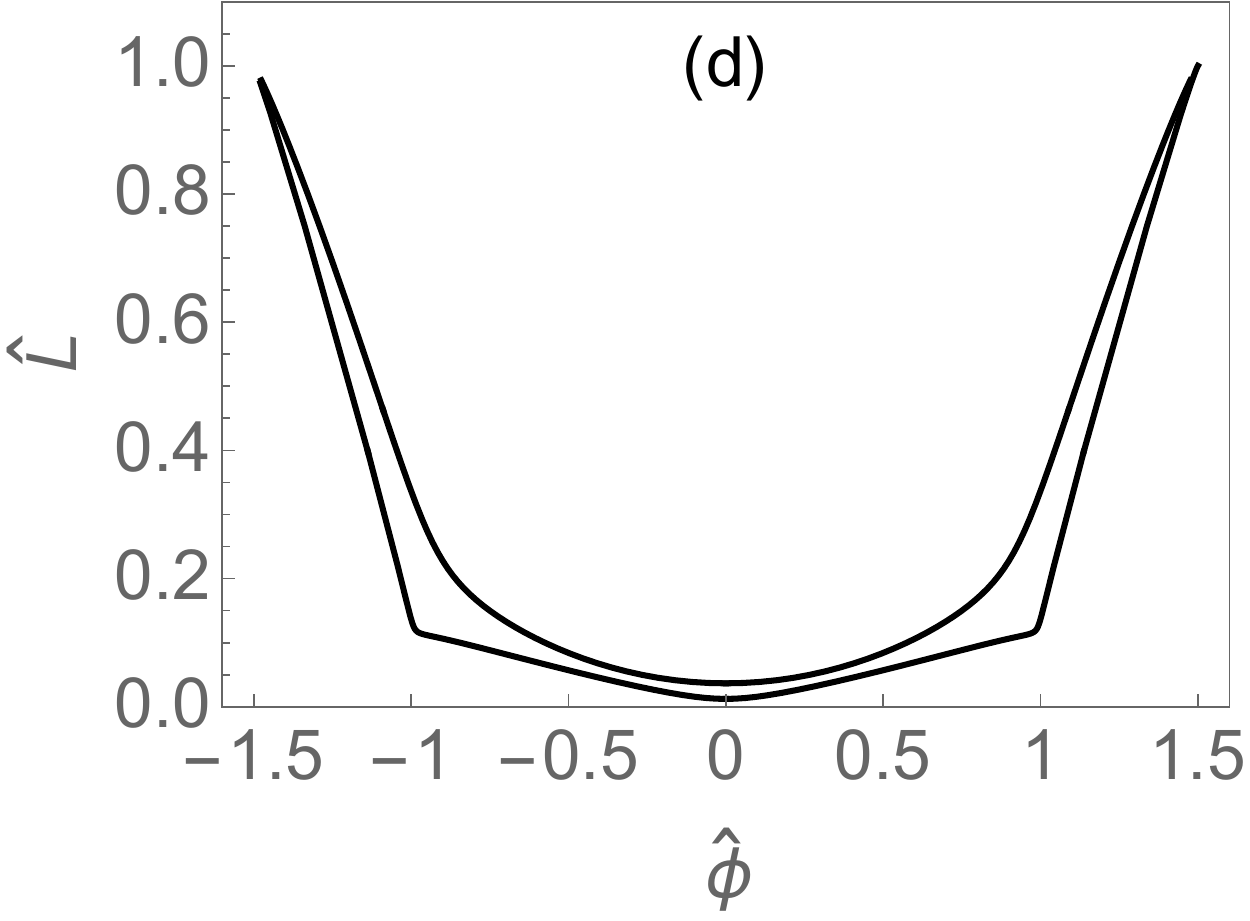}
\end{center}
\caption{(a) Contraction distance $\hat{L}$, (b) twist angle $\hat{\phi}$, and (c) entropy $\hat{S}$, all versus torque $\hat{\tau}$ at constant tension $\hat{J}$ and temperature $\hat{T}$. Panel (d) shows $\hat{L}$ versus $\hat{\phi}$, demonstrating the buckling transition.
The two curves in each panel are for $\hat{T}=1, 0.5$. 
The remaining specifications are from (\ref{eq:30a}), (\ref{eq:30b}).}
  \label{fig:figure26}
\end{figure}

In our methodology, particles from the same species with similar activation densities (implying similar population densities) can describe very different physical phenomena.
The differences are encoded in the specifications (\ref{eq:34}) and visually enhanced in the modified scaling convention (\ref{eq:33}).
The new physics, illustrated in Fig.~\ref{fig:figure26}, includes the buckling transition, of which there is no hint in Fig.~\ref{fig:figure18}.

All it took was that we assigned a smaller contraction length to twist particles. 
This change is conspicuous in the panels (a) and (b) of the two graphs, less so in panel (c).
The buckling transition is manifest in panel (d) in the form of a sharp increase in the rate of contraction.
As is typical of phase behavior in one-dimensional systems, what looks like real transitions are, in fact, crossovers whose width varies with scaled temperature.

The energetics of the relevant particles in the case under scrutiny here is such that fluctuations are sufficiently strong at $\hat{T}=1.0$ to remove any sign of a transition. 
Yet, at $\hat{T}=0.5$, the abrupt initiation of supercoiling is clearly visible in the sharply edged curve of contraction length versus twist angle.
Keep in mind that a change in $\hat{T}$ does not have to mean a change in temperature.

We should like to emphasize that this scenario is not meant to be a model for the buckling transition as observed in ds-DNA.
Nor is it meant to claim predictive power of our methodology in regard to this phenomenon.
What will be necessary to achieve these goals is to fix model parameters with empirical evidence drawn from experimental data unrelated to the buckling transition and then test the supercoil modeling introduced here, appropriately generalized to include native chirality, for its predictive power regarding the buckling transition. 
This strategy worked well in Sec.~V of Ref.~\cite{mct1}) for the force extension characteristics of torsionally unconstrained DNA stretching and will be adapted in the continuation of this project \cite{mct3} to interpret the ubiquitous ``hat curves'' found experimentally in evidence of overwound and underwound DNA buckling \cite{SAB+96, Mark98, ABLC98, SABC98, HYZ99, SAB+99, BM00, ZZY00, SABC00, SACB00, BBS03, Neuk04, Mark07, FDS+08, DFS+09, SW09, MN13, ELS13}.

%
\section{Conclusion and outlook}\label{sec:con-out}
%
In this work we have extended a method of statistical mechanical analysis based on statistically interacting quasi-particles \cite{Hald91a, Wu94, Isak94, Anghel, NA14, LVP+08, copic, picnnn, pichs, LMK09, PMK07, sivp}, previously adapted to molecular molecular chains under tension \cite{mct1}.
The extension includes torsional constraints and adds torque to tension as a second mechanical control variable.
All thermodynamic relations describing quasistatic processes involving combinations of stretching or contracting, twisting and supercoiling are derived from a single partition function.

We have introduced the extended method of analysis in applications to idealized ladder systems and then demonstrated its usefulness for two applications to DNA: (i) the conversion of (native) B-DNA into (underwound) S-DNA or (overwound) P-DNA when subjected to controlled variations of tension and torque, (ii) the conversion between twist chirality and plectonemic chirality when the molecular chain is subjected to controlled variations of linkage and extension.
An arena for applications to the buckling of overwound and underwound ds-DNA has thus been opened up and is one natural extension of this project which we intend to explore \cite{mct3}.

A second projected extension has its focus on the inclusion of interactions between molecular chains and molecules of the embedding fluid (beyond its role as a heat bath).
This extension addresses the effects of intercalation on force-extension characteristics, the role of melting bubbles, and the observed manifestations of hysteresis \cite{PQSC94, RB01, WWRB02, DMZ+02, WR02, VMRW05, PHMV05, SMRW08, WPG08, FCMY10, Zoli11, ZCF+12, BEB+12, ZCL+13, BMLB14, BEB+14, APRW16}. 
Some of these effects can be described within the same framework of statistical mechanical analysis as as anticipated in Sec.~I of Ref.~\cite{mct1}.
However, an extension of this methodology to include kinetics will be necessary and promises to significantly widen its scope \cite{mct3}.




\begin{thebibliography}{100}

\bibitem{MS94}
J. F. Marko and E. D. Siggia,
Fluctuations and supercoiling of DNA,
Science \textbf{265}, 508 (1994).

\bibitem{MS95}
J. F. Marko and E. D. Siggia,
Statistical mechanics of supercoiled DNA,
Phys. Rev. E \textbf{52}, 2912 (1995).

\bibitem{SAB+96}
T. R. Strick, J.-F. Allemand, D. Bensimon, A. Bensimon, V. Croquette, 
The elasticity of a single supercoiled DNA molecule, 
Science \textbf{271}, 1835 (1996).

\bibitem{Mark97}
J. F. Marko,
Stretching must twist DNA,
Europhys. Lett. \textbf{38}, 183 (1997).

\bibitem{Mark98}
J. F. Marko,
DNA under high tension: overstretching, undertwisting, and relaxation dynamics,
Phys. Rev. E \textbf{57}, 2134 (1998).

\bibitem{ABLC98} 
J. F. Allemand, D. Bensimon, R. Lavery, and V. Croquette, 
Stretched and overwound DNA forms a Pauling-like structure with exposed bases.
Proc. Natl. Acad. Sci. USA \textbf{95}, 14152 (1998).

\bibitem{Smit98}
S. B. Smith,
Twisting DNA molecules,
Biophys. J. \textbf{74}, 1609 (1998).

\bibitem{SABC98}
T. R. Strick, J. F. Allemand, D. Bensimon, and V. Croquette,
Behavior of supercoiled DNA,
Biophys. J. \textbf{74}, 2016 (1998).

\bibitem{HYZ99}
Z. Haijun, Z. Yang, and O.-Y. Zhong-can,
Bending and base-stacking interactions in double-stranded DNA,
Phys. Rev. Lett. \textbf{82}, 4560 (1999).

\bibitem{SAB+99}
T. R. Strick, J.-F. Allemand, D. Bensimon, R. Lavery, and V. Croquette,
Phase coexistence in a single DNA molecule,
Physica A \textbf{263}, 392 (1999).

\bibitem{LRS+99}
J. F. L{\'e}ger, G. Romano, A. Sarkar, J. Robert, L. Bourdieu, D. Chatenay, and J. F. Marko,
Structural transitions of a twisted and stretched DNA molecule,
Phys. Rev. Lett. \textbf{83}, 1066 (1999).

\bibitem{BM00}
C. Bouchiat and M. M{\'e}zard,
Elastic rod model of a supercoiled DNA molecule,
Eur. Phys. J. E \textbf{2}, 377 (2000).

\bibitem{ZZY00}
H. Zhou, Y. Zhang, and Z. Ou-Yang,
Elastic property of single double-stranded DNA molecules: theoretical study and comparison with experiments,
Phys. Rev. E \textbf{62}, 1045 (2000).

\bibitem{SABC00}
T. R. Strick, J.-F. Allemand, D. Bensimon, and V. Croquette,
Stress-induced structural transitions in DNA and proteins,
Annu. Biophys. Biomol. Struct. \textbf{29}, 523 (2000).

\bibitem{SACB00}
T. R. Strick, J.-F. Allemand, V. Croquette, and D. Bensimon,
Twisting and stretching single DNA molecules,
Prog. Biophys. Mol. Biol. \textbf{74}, 115 (2000).

\bibitem{SLCM01}
A. Sarkar, J.-F. L{\'e}ger, D. Chatenay, and J. Marko,
Structural transitions in DNA driven by external force and torque,
Phys. Rev. E \textbf{63}, 051903 (2001).

\bibitem{BBS03}
C. Bustamante, Z. Bryant, and B. Smith,
Ten years of tension: single-molecule DNA mechanics,
Nature \textbf{421}, 423 (2003).

\bibitem{BSG+03}
Z. Bryant, M. D. Stone, J. Gore, B. Smith, N. R. Cozzarelli, and C. Bustamante,
Structural transitions and elasticity from torque measurements on DNA,
Nature \textbf{424}, 338 (2003).

\bibitem{Neuk04}
S. Neukirch,
Extracting DNA twist rigidity from experimental supercoiling data,
Phys. Rev. Lett. \textbf{93}, 198107 (2004).

\bibitem{LJL+06}
T. Lionnet, S. Joubaud, R. Lavery, D. Bensimon, and V. Croquette,
Wringing out DNA,
Phys. Rev. Lett. \textbf{96}, 178102 (2006).

\bibitem{Mark07}
J. F. Marko,
Torque and dynamics of linking number relaxation in stretched supercoild DNA,
Phys. Rev. E \textbf{76}, 021926 (2007).

\bibitem{FDS+08}
S. Forth, C. Deufel, M. Y. Sheinin, B. Daniels, J. S. Sethna, and M. D. Wang,
Abrupt buckling transition observed during the plectoneme formation of individual DNA molecules, 
Phys. Rev. Lett. \textbf{100}, 148301 (2008). 

\bibitem{DFS+09}
B. C. Daniels, S. Forth, M. Y. Sheinin, M. D. Wang, and J. P. Sethna,
Discontinuities at the DNA supercoiling transition,
Phys. Rev. E \textbf{80}, 040901 (2009). 

\bibitem{SW09}
M. Y. Sheinin and M. D. Wang,
Twist-stretch coupling and phase transition during DNA supercoiling,
Phys. Chem. Chem. Phys. \textbf{11}, 4800 (2009).

\bibitem{WRC09}
M. C. Williams, I. Rouzina, and M. J. McCauley,
Peeling back the mystery of DNA overstretching,
PNAS \textbf{106}, 18047 (2009).

\bibitem{NR11}
C. Nisoli and A. R. Bishop,
Thermomechanics of DNA: theory of thermal stability under load,
Phys. Rev. Lett. \textbf{107}, 068102 (2011).

\bibitem{GLO+11}
P. Gross, N. Laurens, L. B. Oddershede, U. Bockelmann, E. J. G. Peterman, and G. J. L. Wuite,
Quantifying how DNA stretches, melts and changes twist under tension,
Nature Physics \textbf{7}, 731 (2011).

\bibitem{OK11}
T. Okushima and H. Kuratsuji,
DNA as a one-dimensional chiral material: application to the structural transition between B form and Z form,
Phys. Rev. E \textbf{84}, 021926 (2011).

\bibitem{OK12}
T. Okushima and H. Kuratsuji,
DNA as a one-dimensional chiral material II: dynamics of the structural transition between B form and Z form,
Phys. Rev. E \textbf{86}, 041905 (2012).

\bibitem{MN13}
J. F.Marko and S. Neukirch,
Global force-torque phase diagram for the DNA double helix: structural transitions, triple points, and collapsed plectonemes,
Phys. Rev. E \textbf{88}, 062722 (2013).

\bibitem{ELS13}
M. Emmanuel, G. Lanzani, and H. Schiessel,
Multiplectoneme phase of double-stranded DNA under torsion,
Phys. Rev. E \textbf{88}, 022706 (2013).

\bibitem{mct1}
A. C. Meyer, Y \"Oz, N. Gundlach, M. Karbach, P. Lu, and G. M\"uller, 
Molecular chains under tension:
thermal and mechanical activation of statistically interacting extension and contraction particles, 
Phys. Rev. E \textbf{101}, 022504 (2020).

\bibitem{Hald91a}
F. D. M. Haldane, 
Fractional statistics in arbitrary dimensions: a generalization
of the Pauli principle,
Phys. Rev. Lett. \textbf{67}, 937 (1991).

\bibitem{Wu94}
Y.-S. Wu, 
Statistical distribution for generalized ideal gas of fractional-statistics particles,
Phys. Rev. Lett. \textbf{73}, 922 (1994).

\bibitem{Isak94}
S. B. Isakov, 
Statistical mechanics for a class of quantum statistics,
Phys. Rev. Lett. \textbf{73}, 2150 (1994); 
Generalization of statistics for several species of identical particles,
Mod. Phys. Lett. B \textbf{8}, 319 (1994).

\bibitem{Anghel}
D.-V. Anghel, 
The thermodynamic limit for fractional exclusion statistics,
J. Phys. A \textbf{40}, F1013 (2007);
The fractional exclusion statistics amended,
Europhys. Lett. \textbf{87}, 60009 (2009).

\bibitem{NA14}
G. A. Nemnes and D.-V. Anghel,
Fractional exclusion statistics in non-homogeneous interacting particle systems,
Roman. Rep. Phys. \textbf{66}, 336 (2014).

\bibitem{LVP+08}
P. Lu, J. Vanasse, C. Piecuch, M. Karbach, and G. M{\"u}ller, 
Statistically interacting quasiparticles in Ising chains,
J. Phys. A \textbf{41}, 265003 (2008).

\bibitem{copic}
D. Liu, P. Lu, G. M\"uller, and M. Karbach,
Taxonomy of particles in Ising spin chains,
Phys. Rev. E \textbf{84}, 021136 (2011).

\bibitem{picnnn}
P. Lu, D. Liu, G. M\"uller, and M. Karbach,
Interlinking motifs and entropy landscapes of statistically interacting particles,
Condens. Matter Phys. \textbf{15}, 13001 (2012) [arXiv:1108.2990].

\bibitem{pichs}
D. Liu, J. Vanasse, G. M\"uller, and M. Karbach,
Generalized Pauli principle for particles with distinguishable traits,
Phys. Rev. E \textbf{85}, 011144 (2012).

\bibitem{LMK09}
P. Lu, G. M\"uller, and M. Karbach, 
Quasiparticles in the XXZ model,
Condensed Matter Physics, \textbf{12}, 381 (2009) [arXiv:0909:2728].

\bibitem{PMK07}
G. G. Potter, G. M\"uller, and M. Karbach,
Thermodynamics of ideal quantum gas with fractional statistics in D dimensions,
Phys. Rev. E \textbf{75}, 061120 (2007); 
Thermodynamics of statistically interacting quantum gas in D dimensions,
\textbf{76}, 061112 (2007).

\bibitem{sivp}
B. Bakhti, M. Karbach, P. Maass, M. Mokim, and G. M\"uller,
Statistically interacting vacancy particles,
Phys. Rev. E \textbf{89}, 012137 (2014).

\bibitem{GKLM13}
N. Gundlach, M. Karbach, D. Liu, and G. M\"uller,
Jammed disks in a narrow channel: criticality and ordering tendencies,
J. Stat. Mech. \textbf{P04018} (2013).

\bibitem{janac2}
C. Moore, D. Liu, B. Ballnus, M. Karbach, and G. M\"uller,
Disks in narrow channel jammed by gravity and centrifuge: profiles of 
pressure, mass density, and entropy density,
J. Stat. Mech. \textbf{P04008} (2014).

\bibitem{cohetra}
G. P. Sharma, Y. K. Reshetnyak, O. A. Andreev, M. Karbach, and G. M\"uller,
Coil-helix transition of polypeptide at water-lipid interface,
J. Stat. Phys. \textbf{P01034} (2015).

\bibitem{CLH+96} 
P. Cluzel, A. Lebrun, C. Heller, R. Lavery, J.-L. Viovy, D. Chatenay, and F. Caron,  
DNA: an extensible molecule, 
Science \textbf{271}, 792 (1996).

\bibitem{CYL+04} 
S. Cocco, J. Yan, J.F. Leger, D. Chatenay, and J.F. Marko (2004) 
Overstretching and force-driven strand separation of double-helix DNA, 
Phys. Rev. E \textbf{70}, 011910 (2204).

\bibitem{Wang97} 
M. D. Wang, H. Yin, R. Landick, J. Gelles, and S. M. Block, 
Stretching DNA with optical tweezers, 
Biophysical Journal \textbf{72}, 1334 (1997).

\bibitem{mct3}
A. C. Meyer, M, Karbach, P. Lu, and G. M\"uller,
Environmental effects on the elastic response of molecular chains to tension and torque, (unpublished).

\bibitem{PQSC94}
T. T. Perkins, S. R. Quake, D. E Smith, and S. Chu,
Relaxation of a single DNA molecule observed by optical microscopy,
Science \textbf{264}, 822 (1994).

\bibitem{RB01}
I. Rouzina and V. A. Bloomfield,
Force-induced melting of the DNA double helix: 1. thermodynamic analysis; 2. effect of solution conditions,
Biophys. J. \textbf{80}, 882, 894 (2001).

\bibitem{WWRB02}
J. R. Wenner, M. C. Williams, I. Rouzina, and V. A. Bloomfield,
Salt deendence of the elasticity and overstretching transition of single DNA moldecules,
Biophys. J. \textbf{82}, 3160 (2002).

\bibitem{DMZ+02}
M.-N. Dessinges, B. Maier, Y. Zhang, M.Peliti, D. Bensimon, and V. Croquette,
Stretching single stranded DNA, a model polyelectrolyte,
Phys. Rev. Lett. \textbf{89}, 248102 (2002).

\bibitem{WR02}
M. C. Williams and I. Rouzina,
Force spectroscopy of single DNA and RNA molecules,
Curr. Opin. Struct. Biol. \textbf{12}, 330 (2002).

\bibitem{VMRW05}
I. D. Vladescu, M. J. McCauley, I. Rouzina, and M. C. Williams,
Mapping the phase diagram of single DNA molecule force-induced melting in the presence of ethidium,
Phys. Rev. Lett. \textbf{95}, 158102 (2005).

\bibitem{PHMV05}
O. Punkkinen, P. L. Hansen, and I. Vattulainen,
DNA ovestretching transition: ionic strength effects,
Biophys. J. \textbf{89}, 967 (2005).

\bibitem{SMRW08}
L. Shokri, M. J. McCauley, I. Rouzina, and M. C. Williams,
DNA overstretching in the presence of glyoxal: structural evidence of force-induced DNA melting,
Biophys. J. \textbf{95}, 1248 (2008).

\bibitem{WPG08}
S. Whitelam, S. Pronk, and P. L. Geissler,
There and (slowly) back again: entropy driven hysteresis in a model of DNA overstretching,
Biophys. J. \textbf{94}, 2452 (2008).

\bibitem{FCMY10}
H. Fu, H. Chen, J. F. Marko, and J. Yan,
Two distinct overstretched DNA states,
Nucleic Acids Res. \textbf{38}, 5594 (2010).

\bibitem{Zoli11}
M. Zoli,
Thermodynamics of twisted DNA with solvent interaction,
J. Chem. Phys. \textbf{135}, 115101 (2011).

\bibitem{ZCF+12}
X. Zhang, H. Chen, H. Fu, P. S. Doyle, and J. Yan,
Two distinct overstretched DNA structures revealed by single molecule thermodynamic measurements,
PNAS \textbf{109}, 8103 (2012).

\bibitem{BEB+12}
N. Bosaeus, A. H. El-Sagheer, T. Brown, S. B. Smith, B. Akerman, C. Bustamante, and B. Norden,
Tension induces a base-paired overstretched DNA conformation,
PNAS \textbf{109}, 15179 (2012).

\bibitem{ZCL+13}
G. A. King, P. Gross, U Bockelmann, M. Modesti, G. J. L. Wuite, and E. J. G. Peterman,
Revealing the competition between peeled ssDNA, melting bubbles, and S-DNA during DNA overstretching by single-molecule calorimetry,
PNAS \textbf{110}, 3865 (2013).

\bibitem{BMLB14}
L. Bongini, L. Melli, V. Lombardi, and P. Bianco,
Transient kinetics measured with force steps discriminate between double-stranded DNA elongation and melting and define the reaction energetics,
Nucleic Acids Res. \textbf{42}, 3436 (2014).

\bibitem{BEB+14}
N. Bosaeus, A. H. El-Sagheer,T. Brown, B. Akerman, and B. Norden,
Force-induced melting of DNA -- evidence for peeling and internal melting from force spectra on short synthetic duplex sequences,
Nucleic Acid Res. \textbf{42}, 8083 (2014).

\bibitem{APRW16}
A. A. Almaqwashi, T. Paramanathan, I. Rouzina, and M. C. Williams,
Mechanism of small molecule-DNA interactions probed by single-molecule force spectroscopy,
Nucleic Acids Res. \textbf{44}, 3971 (2016).

















\end{thebibliography}
\end{document}